\documentclass[twocolumn,nofootinbib,superscriptaddress,preprintnumbers,floatfix]{revtex4-1}
\usepackage{amsmath}
\usepackage{amsfonts}
\usepackage{amssymb}
\usepackage{graphicx}
\usepackage{epsfig}
\usepackage{subfigure}
\usepackage{color}
\usepackage{float}
\usepackage{slashed}
\usepackage{hyperref}

\newcommand{\ba}{\begin{align}}
\newcommand{\ea}{\end{align}}
\def\Li{\textrm{Li}}
\def\ln{\textrm{ln}}
\def\df{\textrm{d}}
\def\nn{\nonumber}
\def\MS{\overline{\rm MS}}

\def\LQCD{\Lambda_{\rm QCD}}

\def\bare{{\textrm{bare}}}

\begin{document}

%%%%%%%%%%%%%%%%%%%%%%%%%%%%%%%%%%%%%%%%%%%%%%%%%%%%%%%%%%%%%%%%%%%%%%%%%%%%%%%%
% Title page
%%%%%%%%%%%%%%%%%%%%%%%%%%%%%%%%%%%%%%%%%%%%%%%%%%%%%%%%%%%%%%%%%%%%%%%%%%%%%%%%

\preprint{\vbox{\hbox{UWTHPH 2015-19}}}
\preprint{\vbox{\hbox{DESY 15-098}}}
%\vfill
\title{\Large Variable Flavor Number Scheme for Final State Jets in DIS}

\author{Andre H. Hoang}
\affiliation{Fakult\"at f\"ur Physik, Universit\"at Wien,
Boltzmanngasse 5, 1090 Vienna, Austria}
\affiliation{Erwin Schr\"odinger International Institute for Mathematical Physics, Universit\"at Wien, Boltzmanngasse 9, 1090 Vienna, Austria}

\author{Piotr Pietrulewicz}
\affiliation{Theory Group, Deutsches Elektronen-Synchrotron (DESY),
D-22607 Hamburg, Germany}

\author{Daniel Samitz}
\affiliation{Fakult\"at f\"ur Physik, Universit\"at Wien,
Boltzmanngasse 5, 1090 Vienna, Austria}

\begin{abstract}

We discuss massive quark effects in the endpoint region $x \to 1$ of inclusive deep inelastic scattering (DIS), where the hadronic final state is collimated and thus represents a jet. In this regime heavy quark pairs are generated via secondary radiation, i.e.~due to a gluon splitting in light quark initiated contributions starting at $\mathcal{O}(\alpha_s^2)$ in the fixed-order expansion. Based on the factorization framework for DIS in the endpoint region for massless quarks in Soft Collinear Effective Theory (SCET), which we also scrutinize in this work, we construct a variable flavor number scheme that deals with arbitrary hierarchies between the mass scale and the kinematic scales. The scheme exhibits a continuous behavior between the massless limit for very light quarks and the decoupling limit for very heavy quarks.
It entails threshold matching corrections, arising from all gauge invariant factorization components at the mass scale, which are related to each other via consistency conditions. This is explicitly demonstrated by recalculating the known threshold correction for the parton distribution function at $\mathcal{O}(\alpha_s^2 C_F T_F)$ within SCET. The latter contains large  rapidity logarithms $\sim \ln(1-x)$ that can be summed by exponentiation. Their coefficients are universal, which can be used to obtain potentially relevant higher order results for generic threshold corrections at colliders from computations in DIS. In particular, we extract the $\mathcal{O}(\alpha_s^3)$ threshold correction multiplied by a single rapidity logarithm from results obtained earlier.

\end{abstract}

\maketitle

\section{Introduction}

For multiple collider processes precision phenomenology requires a thorough understanding of quark mass effects to meet future experimental data with high statistics and low systematic uncertainties. One of the challenges hereby is that the quark mass represents a scale that can be parametrically much different than the typical hard scattering scale, which can give rise to large logarithms in fixed order perturbation theory. At the Large Hadron Collider (LHC) this concerns e.g.~bottom mass effects in Higgs production via gluon fusion or in association with bottom quark jets.

One example where the treatment of massive quark effects raised a lot of interest is deep-inelastic scattering (DIS), the benchmark process for the extraction of parton distribution functions (PDFs). These are one main input for the analysis of all processes at hadron colliders, such that precise predictions including the effects of the charm and bottom quark masses are necessary. A first systematic approach to incorporate heavy quarks with arbitrary masses with respect to the other relevant scales has been provided in Refs.~\cite{Aivazis:1993pi,Aivazis:1993kh}, which laid the basis of a {\it variable flavor number scheme} (VFNS) for inclusive processes in hadron collisions. The method is founded on the separation of close-to-mass-shell modes and offshell fluctuations and is thus in the spirit of effective field theory (EFT) factorization, see e.g. Ref.~\cite{Bonvini:2015pxa}. Nowadays, different schemes have been developed to cope with this challenge, which are mainly based on this approach, but they 
differ in their detailed implementation 
concerning formally subleading contributions (see Ref.~\cite{Olness:2008px} for a short overview). 

In this work we discuss the endpoint region of DIS, i.e. $x \to 1$, where the final state becomes a single jet with invariant mass $Q\sqrt{1-x} \ll Q$, where $q^2=-Q^2$ denotes the hard momentum transfer. Although having a limited phenomenological impact, the regime $1-x\ll 1$ provides a simple and instructive example of how to incorporate quark mass effects in differential distributions at hadron colliders with multiple kinematic scales. We construct a VFNS, which is in the same spirit as the well-known VFNS in the classical operator product expansion (OPE) region $1-x \sim \mathcal{O}(1)$~\cite{Aivazis:1993pi,Aivazis:1993kh} and exhibits similar main characteristics. These include (i) the resummation of all large logarithms, also those involving the quark mass, (ii) the correct limiting behavior of the perturbative structures, i.e.~the hard matching coefficient at the scale $Q$ and the jet function at the scale $Q\sqrt{1-x}$, for very small and large masses and (iii) a continuous description for arbitrary 
hierarchies of the 
dynamic scales 
with respect to the mass keeping the full mass dependence of the singular terms (i.e.~at leading order in the power counting). The latter is in particular relevant, since for a single value of the hard momentum transfer $Q$ or the Bjorken variable $x$ the hierarchies can change significantly when scanning over the respective other variable.

We will see that for $x \rightarrow 1$ quark mass effects arise (mainly) via light quark initiated contributions, where the heavy quarks are produced via secondary radiation through the splitting of an additionally emitted virtual gluon starting at $\mathcal{O}(\alpha_s^2)$. Our treatment of these secondary massive quark corrections relies on the setup developed in Refs.~\cite{Gritschacher:2013pha,Pietrulewicz:2014qza}. In these papers a VFNS for event shape distributions, specifically for thrust, in the dijet region for $e^+ e^-$-collisions was constructed in the framework of SCET. It was shown that in a strict EFT interpretation one has to introduce additional degrees of freedom to the existing massless SCET modes, namely collinear and soft mass modes. These adopt the scaling of the corresponding massless modes if the mass is below the typical invariant mass scale of the massless collinear or ultrasoft modes, respectively, but contain in addition fluctuations around the mass-shell which have to be 
considered 
when 
the massive quark is integrated out. The mass mode picture leads to the emergence of different EFTs, which implies the strict guideline of having the massive quark modes either as fluctuating fields or excluded completed (i.e.\ integrated out).
  
In Ref.~\cite{Pietrulewicz:2014qza} also an alternative formulation of the VFNS is described which does not rely on different EFTs, but is only based on the massless factorization theorem with different renormalization conditions for the massive quark corrections to the matrix elements according to the hierarchy between the mass and the other involved scales. This renormalization procedure is similar to the one for the strong coupling in the presence of massive quarks~\cite{Collins:1978wz} and is also the underlying idea for the formulation of the VFNS for DIS in the OPE region~\cite{Collins:1998rz}. The interpretation in terms of different renormalization conditions is convenient since it automatically takes into account any possible power corrections between the hard or jet scale and the mass scale arising in the EFT picture when the hierarchy between the mass scale and the respective kinematic scale is marginal.\footnote{For all possible scale hierarchies covered within our VFNS approach a strict EFT 
factorization can be constructed that agrees with our result up to nonsingular corrections.} Thus it provides a continuous description 
of the cross section for 
arbitrary masses by construction, whereas in a strict EFT picture (which would enforce expansions) the transitions between different hierarchical scenarios would have to be adapted by nonsingular corrections, which concerns in particular the real radiation thresholds. In the following we will therefore discuss the VFNS only in the formulation relying on renormalization conditions.
 
The renormalization conditions with respect to the massive quark corrections which we are going to impose are either the $\MS$ prescription or an on-shell (OS) prescription.\footnote{For the purely massless quark corrections we always use the $\MS$ scheme.} The common use of the $\MS$ prescription has the feature that the $n_l$ massless quarks and the massive flavor both contribute to the renormalization group (RG) evolution in the same way corresponding to an ($n_l+1$) running flavor scheme. The OS prescription is defined by the condition that the massive quark corrections vanish for invariant mass scales much smaller than the quark mass and also subtracts finite and scale-dependent contributions such that the massive flavor does not lead to any contribution in the RG evolution, implying only $n_l$ running flavors. This concerns the matrix elements, i.e.~the current, the jet function and the PDFs, as well as the strong coupling $\alpha_s$. The $\MS$ prescription is appropriate to cover the situation where 
the quark mass becomes 
small (where appropriate means that no large mass logarithms arise in this limit) and leads to expressions which give the known results for massless quarks in the limit $m\rightarrow 0$. The OS prescription is suitable to cover the decoupling limit, such that the effects of the massive quark vanish in the infinite mass limit. 
The differences of the renormalized quantities with respect to both of these renormalization prescriptions constitute matching factors, also called threshold corrections. Since the hard matching coefficient, the jet function and the PDFs are independent and in principle not exclusively tied to any particular factorization theorem, these factors represent also universal ingredients that appear in a similar way for the description of different processes. Here we will emphasize universal features of the threshold corrections and establish the connections to some of the results anticipated in Ref.~\cite{Pietrulewicz:2014qza}.

The outline of this paper is as follows: In Sec.~\ref{sec:massless} we set up the notation and display the massless factorization theorem for the structure functions in DIS for $1-x \ll 1$. Here we do not require to be in the kinematic region with the scaling $1-x \sim \LQCD/Q$ as frequently adopted in the literature. We show that the factorization theorem has the same form in the complete endpoint region $1-x\gtrsim \LQCD/Q$ using the proper mode setup for $1-x \gg\LQCD/Q$. We also explain that massive quark effects can only arise via secondary radiation, and we show in Sec.~\ref{sec:massive} how to incorporate them consistently by setting up a VFNS for any gauge invariant component of the factorization theorem. For definiteness we discuss in Sec.~\ref{sec:applications} practical implementations of the VFNS for various hierarchies between the kinematic scales and the mass scale. Here we also consider different choices of the final renormalization scale in the factorization theorem that lead to 
consistency conditions between the 
threshold correction factors involved in the RG running of the corresponding matrix elements. In Sec.~\ref{sec:PDFmatching} we explicitly calculate the PDF threshold correction at $\mathcal{O}(\alpha_s^2)$ in the large $x$ limit in the effective theory and show that our result is consistent both with the one obtained in classical DIS~\cite{Buza:1995ie} expanded for $x \rightarrow 1$ and with the result for the jet and hard function threshold corrections performed at $\mathcal{O}(\alpha_s^2)$ in Ref.~\cite{Pietrulewicz:2014qza}. In the endpoint region the threshold corrections contain large logarithms $\sim \alpha_s^2 \log$ related to the separation of the collinear and soft mass modes in rapidity, whose resummation we carry out explicitly via the rapidity RGE~\cite{Chiu:2011qc,Chiu:2012ir}. Based on the considerations in Secs.~\ref{sec:applications} and~\ref{sec:PDFmatching} we display also the explicit expressions for the threshold corrections  up to the required order for a full N$^3$LL analysis, which 
includes rapidity logarithms $\sim \alpha_s^2\log$, $\sim \alpha_s^3\log$ and $\sim \alpha_s^4\log^2$. These results represent 
universal ingredients useful for various collider processes 
that involve PDFs in the endpoint region, jet functions and a hard function related to the one appearing in DIS. Finally, in Sec.~\ref{sec:conclusions} we conclude. For the sake of comparison, we provide in Appendix~\ref{sec:fixed-order} the results for secondary massive quark corrections in the OPE region known from Ref.~\cite{Buza:1995ie}. In Appendix~\ref{sec:expansion} we show that for $x \to 1$
our results are in agreement with them.
 
\section{Massless Factorization Theorem for DIS in the Endpoint Region}\label{sec:massless}

 Before discussing quark mass effects we briefly describe the kinematic setup and the factorization theorem for DIS in the endpoint region ${1-x \ll 1}$. Here we display the mode setup, highlight the relevant steps for its derivation specifically for the hierarchy ${1-x \gg \LQCD/Q}$ and show that it can be readily combined with the commonly considered scaling ${1-x \sim \LQCD/Q}$.

\subsection{Kinematics of DIS}
  
 In the following we consider the scattering of an electron off a proton via photon exchange. We denote the proton momentum by $P^\mu$, the momentum of the incoming (outgoing) electron by $k^\mu$ ($k'^\mu$), the incoming momentum of the virtual photon by $q^\mu=k'^\mu-k^\mu$ with spacelike invariant mass $q^2=-Q^2<0$ and the momentum of the outgoing hadronic final state X by $P^\mu_X$. The Lorentz invariant Bjorken scaling variable $x$ is defined by 
 \begin{align}
   x=-\frac{q^2}{2P\cdot q}=\frac{Q^2}{2P\cdot q}  \, 
 \end{align}
 with the kinematic constraint $0 \leq x \leq 1$. We will work in the Breit frame, where $q^\mu$ does not have an energy component and the initial state proton is $\bar{n}$-collinear. Neglecting the proton mass the relevant momenta in the Breit frame in terms of lightcone coordinates read
 \begin{align}\label{eq:PX}
  &q^\mu=\frac{Q}{2}n^\mu - \frac{Q}{2} \bar{n}^\mu \, , \quad P^\mu=\frac{Q}{2x}\bar{n}^\mu, \notag \\& P^\mu_X = \frac{Q}{2}n^\mu + \frac{Q(1-x)}{2x} \bar{n}^\mu \, .
 \end{align}
 In the endpoint region the hadronic final state is an $n$-collinear jet with an invariant mass $P_X^2\approx Q^2(1-x) \ll Q^2$.
 
 The differential cross section for DIS can be decomposed in terms of a leptonic and a hadronic tensor. The latter is defined by
 \begin{align}\label{eq:hadronic_tensor}
  W^{\mu \nu}(P,q)  & =  \frac{1}{2\pi} \, {\rm Im}\left[i \int \df^4z \, e^{iqz} \langle P \vert \, T[J^{\mu \dagger}(z) J^{\nu}(0)] \, \vert P \rangle\right] \, ,
 \end{align}
 with $|P\rangle$ denoting the initial proton state and the current $J^{\mu}(z)= \sum_{q_i} e_{q_i}^2 \, \bar{q}_i\gamma^\mu q_i (z)$ summed over all quark flavors $q_i$ with corresponding electric charges $e_{q_i}$. We will just deal with unpolarized DIS, so that a spin average is always implied. Using current conservation, which implies $q^\mu W_{\mu \nu}=0$, one can decompose the hadronic tensor for the parity conserving vector current into the two structure functions $F_1(x,Q^2)$ and $F_2(x,Q^2)$, 
 \begin{widetext}
\begin{align}
 W^{\mu \nu}(P,q) & =-\left(g_{\mu \nu}-\frac{q^{\mu} q^{\nu}}{q^2}\right) F_1(x,Q)+\frac{1}{P\cdot q}\left(P^{\mu}+\frac{q^\mu}{2x}\right)\left(P^{\nu}+\frac{q^\nu}{2x}\right)F_2(x,Q) \nn \\
 & = -g_\perp^{\mu \nu} F_1(x,Q)+\frac{1}{2x}\left(\frac{n^\mu}{2}+\frac{\bar{n}^\mu}{2}\right)\left(\frac{n^\nu}{2}+\frac{\bar{n}^\nu}{2}\right)F_L(x,Q)\, .
\end{align}
\end{widetext}
with $g_\perp^{\mu \nu}=g^{\mu \nu}-1/2(n^\mu \bar{n}^\nu+ \bar{n}^\mu n^\nu)$. Here the longitudinal structure function $F_L(x,Q)$ reads
\begin{align}\label{eq:F_l}
 F_L(x,Q)=F_2(x,Q)-2x F_1(x,Q) \, ,
\end{align}
in terms of $F_1(x,Q)$ and $F_2(x,Q)$.
These structure functions contain physics at different invariant mass scales and thus must be factorized to resum the corresponding large logarithms.

\subsection{Factorization Setup}\label{sec:factorization_setup}
 
In this section we briefly discuss the derivation of the factorization theorem for inclusive DIS for massless quarks in the endpoint region $1-x \ll 1$ and set up the notation employed for the rest of the paper. The factorization can be performed in a multi-step matching procedure and has been carried out already a number of times~\cite{Manohar:2003vb,Becher:2006mr,Chay:2005rz,Idilbi:2006dg,Chen:2006vd,Fleming:2012kb}. However, here we focus on the proper mode setup in the limit $Q(1-x) \gg \LQCD$ and explain why the factorization theorem adopts the same form as for $Q(1-x)\sim \LQCD$. Although this fact has been already stated in several papers (e.g. Refs.~\cite{Becher:2006mr,Chay:2013zya}), we believe that it is worthwhile to give a short derivation using our mode setup.\footnote{We disagree with the mode setup in Ref.~\cite{Becher:2006mr} which assumes nonperturbative messenger modes for the beam remnants at the invariant mass scale $\LQCD\sqrt{1-x} \ll \LQCD$, while Ref.~\cite{Chay:2013zya} never 
explicitly 
displays the scaling of the modes.}

 \begin{figure}
 \centering
 \includegraphics[width=0.85\linewidth]{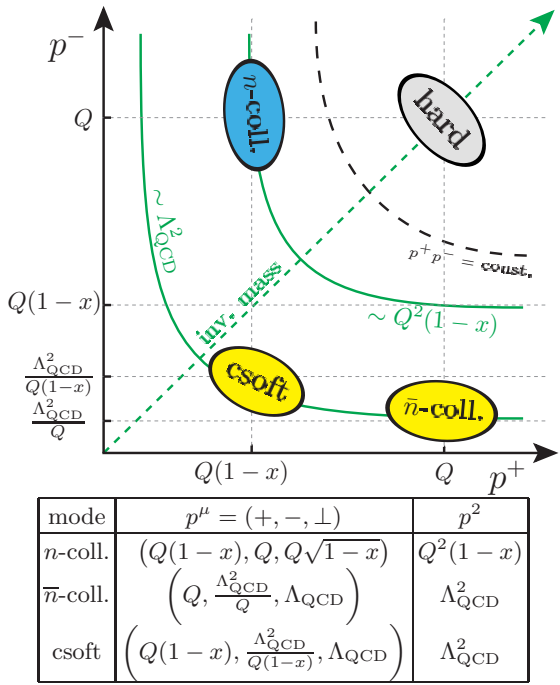}
 \begin{tabular}{|c|c|c|c|}
\hline
mode & $p^\mu=(+,-,\perp)$ & $p^2$\\ 
\hline
$n$-coll. & $\left(Q(1-x),Q,Q\sqrt{1-x}\right)$ & $Q^2 (1-x)$\\
$\overline{n}$-coll. & $\left(Q,\frac{\LQCD^2}{Q},\LQCD \right)$  &$\LQCD^2$\\
csoft & $\left(Q(1-x),\frac{\LQCD^2}{Q(1-x)},\LQCD\right)$ &$\LQCD^2$ \\ %cross-talk between the 2 jets
\hline
\end{tabular}  
\caption{Relevant momentum modes for inclusive DIS in the endpoint region $x \rightarrow 1$ with $1-x \gg \LQCD/Q$. \label{fig:SCET_modes_DIS}}
 \end{figure}
 
The relevant modes are displayed in Fig.~\ref{fig:SCET_modes_DIS}. The $\bar{n}$-collinear modes describing the initial state proton in the Breit frame always have the same scaling $p_{\bar{n}}^\mu =(n\cdot p_{\bar{n}}, \bar{n}\cdot p_{\bar{n}},p_{\bar{n}}^{\perp})\sim (Q,\LQCD^2/Q,\LQCD)$. The final state is strongly collimated for $x\rightarrow 1$ with a large momentum $Q$ and an invariant mass $Q\sqrt{1-x}$ and is thus described by $n$-collinear modes scaling as $p_n^\mu \sim Q(1-x,1,\sqrt{1-x})$. The kinematics in the Breit frame prohibits the appearance of a final $\bar{n}$-collinear state, as can be seen from Eq.~(\ref{eq:PX}). This has the important consequence that the $\bar{n}$-collinear sector just enters the factorization theorem via a component which is local both in label space as well as in the residual coordinate, as has been also pointed out in Ref.~\cite{Fleming:2012kb}. The remaining relevant low-energy modes contribute to the measurement of $x$ or equivalently to the squared invariant mass 
$\sim Q^2(1-x)$ via a component $n\cdot p \sim Q(1-x)$ (i.e.~they have to lie on the vertical line below the $n$-coll.~modes in Fig.~\ref{fig:SCET_modes_DIS}). In fact all such modes give vanishing contributions in perturbation theory, since no physical scale is associated with the other momentum components which results in scaleless integrals. This holds in particular also for ultrasoft modes scaling as $Q(1-x,1-x,1-x)$ as stated e.g.~in Refs.~\cite{Manohar:2003vb,Becher:2006mr,Chay:2013zya}. Thus any additional relevant modes can only be nonperturbative and scale like $p_{cs}^\mu \sim (Q(1-x),\LQCD^2/Q(1-x),\LQCD)$. They encode interference effects between soft initial and final state radiation. Note that in contrast to the case $1-x \sim \LQCD/Q$, where the corresponding modes adopt the soft scaling $p_{s}^\mu \sim \LQCD(1,1,1)$, 
these modes are 
now also boosted in the Breit frame, and 
therefore referred to as collinear-soft (csoft) modes. They are separated by the rapidity factor $(1-
x)$ from the $\bar{n}$-collinear modes. 
These types of modes have recently received some attention in the context of multidifferential cross sections and have been incorporated systematically into a modified version of SCET, called SCET$_+$~\cite{Bauer:2011uc,Procura:2014cba}. We will discuss here the DIS factorization theorem in the same spirit using a multistage matching procedure. However, our case is simpler, since no relevant softer mode is present with which the csoft mode can potentially interact.\footnote{Here only the separation in rapidity matters, i.e.~the "softness" of the csoft mode with respect to the $\bar{n}$-collinear mode, while the boost of the csoft mode in the Breit frame is actually irrelevant.}

\begin{figure*}
  \centering
  \includegraphics[width=0.65\textwidth]{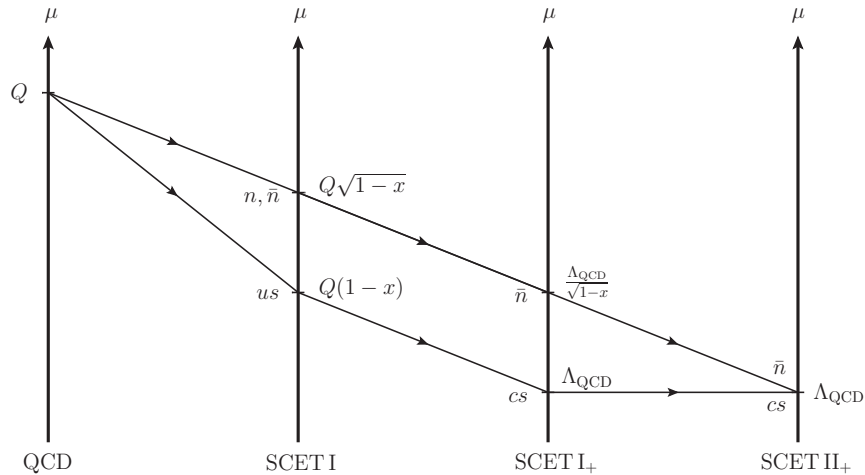}
  \caption{Schematic picture of the multistage matching procedure for $1-x \gg \LQCD/Q$ described in the text. \label{fig:DIS_matching}}
   \end{figure*}
   
To derive the factorization theorem we employ the multistep matching procedure sketched in Fig.~\ref{fig:DIS_matching}. We emphasize that other ways to order some of the matching steps are possible and one may even work in a framework containing the csoft modes from the beginning. Here we first match the QCD current to the usual SCET I current
 \begin{align}\label{eq:SCET_current}
  J_{\rm QCD}^\mu \longrightarrow  J^\mu_{\rm SCET\, I}= C(Q,\mu) \,\bar{\chi}_{n} Y^\dagger_{n} \gamma^\mu Y_{\bar{n}} \chi_{\bar{n}}  \, ,
 \end{align}
 with $\chi_{\bar{n}}\equiv W_{\bar{n}}^\dagger \xi_{\bar{n}}$ and $\bar{\chi}_{n}\equiv \bar{\xi}_n W_n$, where $W_n$ and $W_{\bar{n}}$ denote collinear Wilson lines. The ultrasoft Wilson lines $Y_n$ and $Y_{\bar{n}}$ appear in the current after the BPS field redefinition~\cite{Bauer:2001yt} that disentangles the collinear and ultrasoft sectors in the Lagrangian. SCET I describes collinear fluctuations at the invariant mass scale $Q\sqrt{1-x}$ with residual momenta of order $Q(1-x)$. When lowering the virtualities the $n$-collinear final state modes need to be integrated out, and the ultrasoft sector is being resolved with momentum components $\sim Q(1-x)$ becoming labels. We call the corresponding theory with csoft modes of virtuality $\gtrsim \LQCD$ and $\bar{n}$-collinear modes of virtuality $\gtrsim \LQCD/\sqrt{1-x} \gg \LQCD$ SCET I$_+$.\footnote{We do not consider ultrasoft modes in SCET I$_+$ since they do 
 not 
 contribute to any measurement as stated before.}  Here $\LQCD/\sqrt{1-x}$ is the invariant mass scale of the collinear modes at which they can still interact with the csoft modes via the momentum component $\bar{n}\cdot p_{\bar{n}}\sim \LQCD^2/\big(Q(1-x)\big)$. The matching coefficient between the two theories is the quark jet function $J(s,\mu)$, a vacuum correlator of the hard collinear fields in SCET I describing the production rate of an inclusive jet with invariant mass $s$. It is defined in terms of the $n$-collinear fields as
\begin{align}
 & J(Q r_n^+,\mu)\equiv \frac{-1}{2\pi N_c Q} \, \nn\\& \times \mathrm{Im}\left[ i \int {d^4z\;e^{i r_n \cdot z}}\langle 0|\mathrm{T}\left\{\bar{\chi}_{n,Q}(0)\frac{\slashed{\bar{n}}}{2}\chi_n(z)\right\}|0 \rangle\right] \,, 
\label{eq:jetfunctiondefinition}
\end{align}
where the invariant mass is $r_n^2 \simeq Q r_n^+$ and $\chi_{n,Q} \equiv \delta_{\bar{n} \cdot \mathcal{P},Q} \, \chi_n $ with $\mathcal{P}^\mu$ denoting the label momentum operator. All color and spin indices are traced implicitly. Here and in the following all expressions are only given for initial state quarks. For antiquarks the corresponding expressions are completely analogous. The matrix element in SCET I$_+$ is the quark PDF given by
\begin{align}
\phi_{q/P}\Big(\frac{\ell}{Q}, \mu\Big)=\langle P \vert \bar{\chi}_{\bar{n}} X^\dagger_{\bar{n}} V_{\bar{n}} \frac{\slashed{n}}{2}  \delta(\ell-n\cdot \hat{p}) V_{\bar{n}}^\dagger X_{\bar{n}} \chi_{\bar{n},Q}\vert  P\rangle \, .
\end{align}
Here $V_{\bar{n}}$ and $X_{\bar{n}}$ are Wilson lines of the label and small component of the csoft fields required by gauge invariance, written in the notation of Refs.~\cite{Bauer:2011uc,Procura:2014cba}.\footnote{In Refs.~\cite{Bauer:2011uc,Procura:2014cba} these were obtained by resolving the collinear sector. Integrating out offshell fluctuations of virtuality $\sim Q^2(1-x)$ generate $V_{\bar{n}}$ (in analogy to the hard collinear Wilson line $W_{\bar{n}}$) and the Bauer-Pirjol-Stewart (BPS) field redefinition leads to the emergence of $X_{\bar{n}}$ (in analogy to the ultrasoft Wilson line $Y_{\bar{n}}$).} In our matching procedure they originate directly from boosting the ultrasoft Wilson lines in SCET I
\begin{align}\label{eq:Vn}
Y_n^\dagger & \rightarrow V^\dagger_{\bar{n}}={\rm P} \, {\rm exp}\left[ig \int_0^\infty \df s \, n \cdot A_{cs}(s n^\mu+x^\mu)\right]\, , \\
Y_{\bar{n}} & \rightarrow X_{\bar{n}}={\rm P} \, {\rm exp}\left[ig \int_{-\infty}^0 \df s \, \bar{n} \cdot A_{cs}(s \bar{n}^\mu+x^\mu)\right] \, . \label{eq:Xn}
\end{align}
The change of the modes in the soft sector involves in principle an additional matching coefficient. However, since the ultrasoft sector in SCET I and the csoft sector in SCET I$_+$ are fully decoupled from any other sector and this matching only involves a single sector in both theories, for which we can use the QCD Lagrangian, the form of Eqs.~(\ref{eq:Vn}) and~(\ref{eq:Xn}) makes clear that no nontrivial matching coefficient is generated. As a last step the fluctuations at the invariant mass scale $\LQCD/\sqrt{1-x}$ need to be integrated out to describe the physics at the scale $\sim \LQCD$, the invariant mass of the initial state proton, in terms of the final EFT, which we call SCET II$_+$. Since the interactions between the $\bar{n}$-collinear and csoft sectors are already fully disentangled there is no matching coefficient originating from the collinear sector and the transition from SCET I$_+$ to SCET II$_+$, where these two types of modes cannot interact with each other, and can be achieved 
simply by lowering the virtuality of the $\bar{n}$-collinear modes, in 
analogy to the 
matching between SCET I and SCET II~\cite{Bauer:2002aj}. The PDF in SCET II$_+$ can be written as\footnote{Here the convolution is in fact spurious due to the overall delta-distribution in the collinear function $g_{q/P}$ in Eq.~\eqref{eq:f_qp}, such that the PDF $\phi_{q/P}$ could be also written as a simple product. However, the form in Eq.~\eqref{eq:pdf_x1}  will be more convenient for discussing explicit results since the two functions $S_c$ and $g_{q/P}$ have the same dimension.}
\begin{align}\label{eq:pdf_x1}
\phi_{q/P}(1-z, \mu)= Q\int \mathrm{d}\ell\,g_{q/P}(Q(1-z)-\ell,\mu) \, S_c(\ell,\mu) \, ,
\end{align}
where $g_{q/P}(\ell,\mu)$ denotes a local collinear matrix element
\begin{align}\label{eq:f_qp}
 g_{q/P}(\ell,\mu)= \frac{1}{Q}\langle P \vert \bar{\chi}_{\bar{n}}(0) \frac{\slashed{n}}{2} \chi_{\bar{n},Q}(0) \vert  P\rangle \, \delta(\ell) \, ,
\end{align}
and $S_c(\ell,\mu)$ denotes a vacuum expectation value in terms of csoft fields, the csoft function
\begin{align}
S_c(\ell,\mu)=\frac{1}{N_c}  \sum_{X_{cs}} \langle 0 \vert X^\dagger_{\bar{n}} V_{\bar{n}} (0) \,\delta(\ell-n\cdot \hat{p})\, V_{\bar{n}}^\dagger X_{\bar{n}}(0)\lvert 0\rangle \, , \label{eq:csoftfunction}
\end{align}
where again all color indices are traced implicity. We will see explicitly that both $g_{i/P}(\ell,\mu)$ and $S_c(\ell,\mu)$ individually contain rapidity divergences which cancel in the total PDF in Eq.~(\ref{eq:pdf_x1}). The appearance of the csoft function in Eq.~\eqref{eq:csoftfunction} is the only deviation with respect to the case $1-x \sim \LQCD/Q$, where instead the analogue matrix element with soft fields appears, which reads
\begin{align}\label{eq:S_DIS}
 S(\ell,\mu)=\frac{1}{N_c}  \langle 0 \vert S^\dagger_{\bar{n}} S_{n} (0) \,\delta(\ell-n\cdot \hat{p})\, S^\dagger_{n} S_{\bar{n}}(0)\lvert 0\rangle \, .
\end{align}
Note that the structure of the Wilson lines in $S_c(\ell,\mu)$ and $S(\ell,\mu)$ is identical ($X_{\bar{n}}\leftrightarrow S_{\bar{n}} $, $V^\dagger_{\bar{n}}\leftrightarrow S^\dagger_{n} $ only related by a common boost) and the lack of additional relevant softer modes in SCET II$_+$ implies that the Lagrangian in the csoft sector can be replaced by the full QCD Lagrangian. Thus the interactions for csoft and soft modes are equivalent. So $S_c(\ell,\mu)$ and $S(\ell,\mu)$ give the same result (which we will demonstrate explicitly in Sec.~\ref{sec:one_loop}) and we will not distinguish them any more in the following discussion of the factorization theorems. Since this concerns the only potential difference for the two scaling hierarchies $1-x \gg \LQCD/Q$ and $1-x \sim \LQCD/Q$, we obtain that the factorization theorem in the complete endpoint region is always the same for any $1-x \gtrsim \LQCD/Q$.

We emphasize that in contrast to the OPE regime $1-x \sim \mathcal{O}(1)$ the endpoint PDF does not encode only collinear initial state radiation, but also (c)soft interference effects between initial and final state radiation. Therefore, the PDF at the endpoint may not be interpreted only as a description of the momentum distribution inside the proton before the hard interaction.

The full factorization theorem reads (to all orders in $\alpha_s$ and at leading order in $1-x$)
\begin{align}\label{eq:fact_theorem_x1}
&F_1(x,Q) = \frac{1}{2x}F_2(x,Q)\notag = \, \sum\limits_{i=q,\bar{q}} \frac{e_i^2}{2} \, H^{(n_f)}(Q,\mu)\nn\\&\quad\quad\times\,  \int \df s  \, J^{(n_f)} \left(s,\mu\right) \, \phi^{(n_f)}_{i/P}\left(1-x-\frac{s}{Q^2}, \mu\right) \, ,
\end{align}
where the superscript $(n_f)$ indicates the number of active quark flavors relevant for the RG evolution of all renormalized structures including in particular also the strong coupling constant. Here the hard function $H^{(n_f)}(Q,\mu)$ is the square of the matching coefficient between the SCET and the QCD currents $C^{(n_f)}(Q,\mu)$ in Eq.~\eqref{eq:SCET_current}, while the jet function $J^{(n_f)}(s,\mu)$ and the PDF $\phi^{(n_f)}_{i/P}(\ell,\mu)$ are defined in Eq.~\eqref{eq:jetfunctiondefinition} and Eq.~\eqref{eq:pdf_x1} respectively. 
Note that the hadronic tensor becomes transverse in the limit $x\rightarrow 1$, such that $F_L(x,Q)=0$ and the Callan-Gross relation $F_2(x,Q)=2x F_1(x,Q)$ is satisfied to all orders in $\alpha_s$.

The massless fixed-order hard and jet functions, $H^{(n_f)}(Q,\mu_H)$ and $J^{(n_f)}(s,\mu_J)$, are known up to $\mathcal{O}(\alpha_s^3)$ and  $\mathcal{O}(\alpha_s^2)$, respectively, and the anomalous dimensions are known up to $\mathcal{O}(\alpha_s^3)$. Explicit expressions can be found e.g.~in Ref.~\cite{Becher:2006mr}.
For the hard function we write
\begin{align}\label{eq:groupstructure}
 &H^{(n_{\!f})}(Q,\mu)= \, 1 + H^{(n_{\!f},1)}(Q,\mu)+\Big[H^{(n_{\!f},2)}_{C_F}(Q,\mu) \nn\\&\quad+  H^{(n_{\!f},2)}_{C_A}(Q,\mu) + n_f \, H^{(n_{\!f},2)}_{T_F}(Q,\mu)\Big] +\mathcal{O}(\alpha_s^3)  \, ,
\end{align}
where $H^{(n_{\!f},1)}$, $H^{(n_{\!f},2)}_{ C_F}$, $H^{(n_{\!f},2)}_{C_A}$ and $H^{(n_{\!f},2)}_{T_F}$ denote the contributions at $\mathcal{O}(\alpha_s)$, $\mathcal{O}(\alpha_s^2 C_F^2)$, $\mathcal{O}(\alpha_s^2 C_F C_A)$ and $\mathcal{O}(\alpha_s^2 C_F T_F)$, respectively. We use an analogous notation for all other perturbative expressions throughout this paper. The additional dependence on a finite quark mass will be indicated in the arguments.

The factorization theorem of Eq.~\eqref{eq:fact_theorem_x1} is written with all its components at the common renormalization scale $\mu$, which can be chosen independently from the respective characteristic scales $\mu_H\sim Q$ for the hard function, $\mu_J\sim Q\sqrt{1-x}$ for the jet function and $\mu_\phi\sim \LQCD$ for the PDF. Since the choice of $\mu$ necessarily differs widely from at least two of the characteristic scales, it is mandatory to sum large logarithmic terms using the RG equations. This is achieved by writing each component of the factorization theorem as a function that is defined at the respective characteristic scale $\mu_H$, $\mu_J$ or $\mu_\phi$ supplemented by a RG evolution factor that sums the logarithms between the characteristic scales and the common scale $\mu$:\footnote{
Note the convention concerning the ordering of the arguments of the evolution factors for the hard function in comparison to the jet and PDF functions.
}
\begin{align}
   &H^{(n_f)}(Q,\mu) \nn\\&\quad=  \, H^{(n_f)}(Q,\mu_H) \, U_H^{(n_f)}(Q,\mu_H,\mu) \, , \\
   &J^{(n_f)}(s,\mu)\nn\\&\quad=   \int \df s' \, J^{(n_f)}(s-s',\mu_J) \, U_J^{(n_f)}(s',\mu,\mu_J) \, , \\
   &\phi^{(n_f)}(1-z,\mu)\nn\\&\quad=   \int \df z' \, \phi^{(n_f)}(z'-z,\mu_\phi) \, U_\phi^{(n_f)}(1-z',\mu,\mu_\phi) \, .\label{eq:PDFevolution}
 \end{align}
The individual functions at the respective characteristic scales $\mu_H$, $\mu_J$ and $\mu_{\phi}$, which are free of any large logarithmic terms, serve as the initial conditions of the respective RG evolution which follows the RG equations
\begin{align}
\label{eq:currentRGE_massless}
 &\mu\frac{\df}{\df\mu}\,U^{(n_{\!f})}_{H}(Q,\mu_H,\mu) \nn\\&\quad=\gamma^{(n_{\!f})}_{H}(Q,\mu) \, U^{(n_{\!f})}_{H}(Q,\mu_H,\mu) \, , \\
\label{eq:jetRGE_massless}
 &\mu\frac{\df}{\df\mu}\,U^{(n_{\!f})}_{J}(s,\mu,\mu_J) \nn\\&\quad=\int\! \df s' \,\gamma^{(n_{\!f})}_J(s-s',\mu)\,U^{(n_{\!f})}_{J}(s',\mu,\mu_J) \, , \\
\label{eq:pdfRGE_massless}
 &\mu\frac{\df}{\df\mu}\,U^{(n_{\!f})}_{\phi}(1-z,\mu,\mu_\phi) \nn\\&\quad=\int\! \df z'\, \gamma^{(n_{\!f})}_{\phi}(z'-z,\mu)\,U^{(n_{\!f})}_{\phi}(1-z',\mu,\mu_\phi) \, .
\end{align}
The superscript $(n_f)$ for all components of the factorization theorem (including RG factors) is a reminder that a renormalization scheme with $n_f$ dynamic running quark flavors is used, associated to an $n_f$-flavor scheme. For the hard and jet functions and the PDF this scheme is implemented through the common $\overline{\text{MS}}$ subtraction scheme for all corrections coming from $n_f$ quarks. Subsequently this scheme implies that all these quarks enter the RG equations via a global $n_f$-dependence. We recall that the anomalous dimensions can be determined from the counterterm factors $Z$ that arise in the renormalization procedure of the individual functions. For massless quarks they are defined in the $\overline{\text{MS}}$ scheme. For example, for the PDF one has
\begin{align}
 \gamma_{\phi} (1-z,\mu)& = - \int \df z' \,Z_{\phi}^{-1}(z'-z,\mu) \, \mu \frac{\df}{\df\mu}\, Z_{\phi} (1-z',\mu)\,.
\end{align}

In Eq.~(\ref{eq:fact_theorem_x1}) the choice of $\mu$ is arbitrary, and the dependence on $\mu$ cancels exactly working
to any given order in perturbation theory. The fact that any other choice for $\mu$ can be implemented leads to a consistency relation 
between the renormalization group factors, which reads~\cite{Becher:2006mr}
\begin{align}\label{eq:consistency_ML}
 &Q^2 \,U_{H}^{(n_{\!f})}(Q,\mu_0,\mu)\, U_J^{(n_{\!f})}(Q^2(1-z),\mu,\mu_0) \nn\\&= U_{\phi}^{(n_{\!f})}(1-z,\mu_0,\mu) \, ,
\end{align}
and a corresponding relation for the anomalous dimensions.

Accounting for massive quarks the factorization theorem (\ref{eq:fact_theorem_x1}) stays valid with some modifications. This concerns the mass dependence of the hard and jet functions, which will be indicated in the arguments, and a modified RG evolution with an adapted flavor number according to the hierarchy between the renormalization scale $\mu$ and the mass scale $\mu_m \sim m$, as described below. The consistency relation (\ref{eq:consistency_ML}) remains intact since the UV divergences are mass independent. However, additional consistency relations emerge between threshold corrections arising when massive quark modes are integrated out. In the following sections we will discuss these points in detail.

 \begin{figure}
  \centering
  \includegraphics[width=0.45\textwidth]{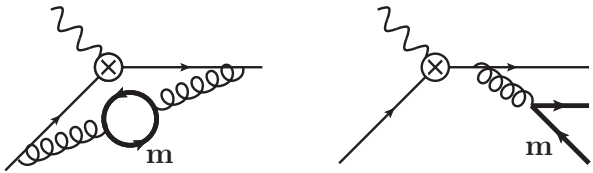}
  \caption{Exemplary diagrams for secondary massive quark production in DIS at $\mathcal{O}(\alpha_s^2 C_F T_F)$. \label{fig:DIS_secondary}}
   \end{figure}
 
An important feature of the factorization theorem in Eq.~(\ref{eq:fact_theorem_x1}) is that there are no flavor mixing terms between quarks and gluons in any of the EFT contributions in the hard current matching, the jet function, the PDF and their evolution factors. One can easily explain this using the possible interactions in SCET at leading order in the power counting parameter. For example, for the PDF evolution one can make the following argument: Flavor mixing requires the splitting of an initial state collinear quark or gluon into two partons. To stay in the endpoint regime one of the final partons has to carry the longitudinal momentum fraction $\xi>x\approx 1$, i.e.~almost the whole momentum, implying that the remaining parton is (ultra)soft. The only available interaction of this kind at leading order in SCET is the emission of an (ultra)soft gluon from a collinear quark, whereas the splitting of a collinear gluon into a collinear and (ultra)soft quark is suppressed by $\mathcal{O}(1-x)$. This 
means 
that the parton extracted out of the PDF at the low scale $\sim \LQCD$ is also the one interacting with the hard photon and entering the final state jet, and thus cannot be a gluon. Since we assume $m\gg \LQCD$, so that the heavy quarks are not produced nonperturbatively out of the proton, this has the consequence that massive quarks enter the EFT components of the factorization theorem only via secondary corrections, i.e.~via contributions which are initiated by massless quarks and where massive quarks are produced through the radiation of virtual gluons that split into a massive quark-antiquark pair, see Fig.~\ref{fig:DIS_secondary}. We mention that in the full QCD current there are also flavor mixing corrections with massive quarks which in general start contributing also at $\mathcal{O}(\alpha_s^2)$ like the secondary corrections. Since these types of corrections do not have a corresponding EFT counterpart, they can be easily included in the hard matching coefficient, and we will not consider them 
specifically in our discussion. In fact, due to Furry's theorem, for the case of the electromagnetic vector current these effects do not show up at $\mathcal{O}(\alpha_s^2 C_F T_F)$ relevant for N$^3$LL resummation, which is the order where we give explicit results.

\section{Massive quark corrections for all components}\label{sec:massive}

In this section we summarize all ingredients of the VFNS for secondary massive quark effects for the most singular terms in inclusive DIS in the endpoint region $x\to 1$. We consider a setup with $n_l$ massless flavors and one massive quark species with mass $m \gg \Lambda_{\text{QCD}}$, which we want to incorporate into the factorization theorem of Eq.~(\ref{eq:fact_theorem_x1}).\footnote{We remark that we will not consider the possibility of having an 
 intrinsic charm contribution with $m \sim \Lambda_{\rm QCD}$. In this case the mass effects in the perturbative corrections are anyway power suppressed.}  This can be easily generalized to the case of several massive quark flavors with different masses appearing in practical considerations where the masses of both the charm and bottom quarks may be relevant. We do not impose any restriction concerning the relation of the mass $m$ to any of the hard, jet, or PDF scales. The massive quark flavor will never be integrated out (in a strict EFT sense) and is thus contained in the full QCD description (relevant for the matching computation for the hard function) as well as in the collinear and soft/usoft sectors of SCET. Compared to Eq.~\eqref{eq:fact_theorem_x1} the factorization theorem for the form factors $F_{1,2}$ has an additional dependence on the mass $m$:
 \begin{align}\label{eq:facttheomassive}
&F_1(x,Q,m) = \frac{1}{2x}F_2(x,Q,m)\notag \\&= \, \sum\limits_{i=q,\bar{q}} \frac{e_i^2}{2} \, H^{(n_f)}(Q,m,\mu)  \int \df s  \, J^{(n_f)} \left(s,m,\mu\right)\nn \\&\quad\quad\times \, \phi^{(n_f)}_{i/P}\left(1-x-\frac{s}{Q^2}, m,\mu\right) \, .
\end{align}
The factorization theorem has an additional residual dependence on the flavor threshold matching scale $\mu_m$. The scale $\mu_m$ is chosen close to the quark mass $m$, $\mu_m \sim m$, but otherwise it is arbitrary and represents the scale at which we switch between the $(n_l)$ and the $(n_l+1)$ running flavor schemes. To be specific, we employ for all components of the factorization scheme depending on the relation of the common scale $\mu$ with respect to $\mu_m$:
\begin{itemize}
 \item the $(n_f)=(n_l)$ flavor scheme for $\mu < \mu_m$ and
 \item the $(n_f)=(n_l+1)$ flavor scheme for $\mu>\mu_m$.
\end{itemize}
These two flavor schemes are implemented independently for each of the components in Eq.~\eqref{eq:facttheomassive} by either using an OS or the $\overline{\text{MS}}$ subtraction prescription for UV-divergent secondary massive quark loop corrections.

In addition, there are flavor threshold matching conditions arising from the difference of the subtraction prescriptions of the two schemes whenever the RG evolution requires a transition through $\mu_m$. The overall RG-invariance of the factorization theorem concerning changes of the renormalization scales and the consistency of properly employing OS and $\overline{\text{MS}}$ subtraction for the secondary massive quark corrections ensure that the factorization theorem is continuous at the RG transition through the threshold scale $\mu_m$.

\subsection{VFNS for the Hard Function}

The hard function $H^{(n_f)}(Q,m,\mu)$ appearing in the factorization theorem of Eq.~\eqref{eq:facttheomassive} is in the $(n_f)=(n_l+1)$ flavor scheme if $\mu$ is above $\mu_m$ and in the $(n_f)=(n_l)$ flavor scheme if $\mu$ is below $\mu_m$:
\begin{align}
   H^{(n_f)}(Q,m,\mu)=\left\{\begin{array}{ll} 
                      H^{(n_l+1)}(Q,m,\mu) \hspace{0.0cm} & {\rm for} \,\, \mu > \mu_m \, ,\\
                      \vspace{-0.2cm} \\
                       H^{(n_l)}(Q,m,\mu) \hspace{0.0cm} & {\rm for} \,\, \mu < \mu_m \, .
                       \label{eq:hardfunction_at_mu}
                     \end{array} \right. 
   \end{align}
The common renormalization scale $\mu$ is in general different from the characteristic scale $\mu_H \sim Q$ of the hard function, so we specify the hard function at the common scale $\mu$ by the hard function at the scale $\mu_H$, which is free of any large logarithmic terms, multiplying a RG-evolution factor that resums the logarithms between the scales $\mu_H$ and $\mu$. The hard function at the scale $\mu_H$ serves as the initial condition of this RG evolution, and the flavor scheme that is employed for the initial condition depends again on the relation of $\mu_H$ to the flavor matching scale $\mu_m$:
 \begin{align}
   H^{(n_f)}(Q,m,\mu_H)=\left\{\begin{array}{ll} 
                      H^{(n_l+1)}(Q,m,\mu_H)   \hspace{0.0cm} & {\rm for} \,\, \mu_H > \mu_m \, ,\\
                      \vspace{-0.2cm} \\
                      H^{(n_l)}(Q,m,\mu_H) \hspace{0.0cm} & {\rm for} \,\, \mu_H < \mu_m \, , \label{eq:hardfunction_at_muH}
                     \end{array} \right. 
   \end{align}
where   
\begin{align}
 H^{(n_l)}(Q,m,\mu_H)= & \, H^{(n_l)}(Q,\mu_H)+2 \hat{F}^{(n_l,2)}_{m} (Q,m) \, , \label{eq:hardcoeff}\\
 H^{(n_l+1)}(Q,m,\mu_H)= & \, H^{(n_l+1)}(Q,\mu_H)+ 2 \hat{F}_{\Delta m}^{(n_l+1,2)} (Q,m) \, .
 \label{eq:hardcoeffI}
 \end{align}
 The functions $H^{(n_l)}(Q,\mu_H)$ and $H^{(n_l+1)}(Q,\mu_H)$ are the hard functions for massless quarks in the $(n_l)$ and $(n_l+1)$ flavor scheme respectively in the notation of Eq.~\eqref{eq:groupstructure}. The term $\hat{F}^{(n_l,2)}_{m}$ represents the massive quark loop contribution to the QCD current form factor with OS subtraction in the $(n_l)$ scheme, see the first diagram in Fig.~\ref{fig:DIS_secondary}. In fact, for the case of DIS in the endpoint region all mass dependent corrections at $\mathcal{O}(\alpha_s^2 C_F T_F)$ can be directly inferred from the matching calculations carried out for thrust in Ref.~\cite{Pietrulewicz:2014qza} due to the fact that the hard coefficients are the same up to an analytic continuation from the timelike to the spacelike process. We write the result as ($\alpha_s^{(n_l)}=\alpha_s^{(n_l)}(\mu_H)$)
\begin{align}\label{eq:F_QCD}
 \hat{F}^{(n_l,2)}_{m}(Q,m)= \frac{\big(\alpha_s^{(n_l)}\big)^2 C_{\!F} T_F}{16\pi^2} \, f^{(2)}_{m}(Q,m)
 \end{align}
 with the function $f^{(2)}_{m}(Q,m)$ given by
 \begin{align}\label{eq:f_QCD}
  &f^{(2)}_{m}(Q,m) =  \, \left\{\Big(\frac{46}{9}r^3+\frac{10}{3}r\Big) \bigg[\Li_2\bigg(\frac{r-1}{r+1}\bigg) -\Li_2\bigg(\frac{r+1}{r-1}\bigg) \bigg]  \right. \nn\\ &\left. +\Big(\!-r^4 + 2 r^2+\frac{5}{3}\Big)\!\bigg[\Li_3\bigg(\frac{r+1}{r-1}\bigg)+\Li_3\bigg(\frac{r-1}{r+1}\bigg) -2\zeta_3\bigg]\right.\nn \\ &\left. +\!\left(\frac{110}{9}r^2+\frac{200}{27}\right)\ln\Big(\frac{1-r^2}{4}\Big) + \frac{238}{9}r^2+\frac{1213}{81} \right\}  ,
\end{align}
where 
\begin{align}
 r=\sqrt{1-4 \hat{m}^2} \, ,
\end{align}
and $\hat{m}=m/Q$.
Due to the imposed OS renormalization condition this correction decouples for large masses, i.e.~$\hat{F}^{(n_l,2)}_{m}(Q,m) \rightarrow 0$ for $\hat{m}\equiv m/Q \to \infty$. For small masses the function $f^{(2)}_{m}(Q,m)$ reads
\begin{align} \label{eq:F_QCD0}
&f^{(2)}_{m}(Q,m) \stackrel{\hat{m} \rightarrow 0}{\longrightarrow}  \, \frac{4}{9}\,\ln^3{(\hat{m}^2)}+\frac{38}{9}\,\ln^2{(\hat{m}^2)} \nn \\&+\bigg(\frac{530}{27}+\frac{4\pi^2}{9}\bigg)\ln{(\hat{m}^2)}+\frac{3355}{81} +\frac{38\pi^2}{27}-\frac{16}{3}\zeta_3 \,  ,
\end{align}
which exhibits mass singularities and is therefore not suitable in the small mass regime. The term $\hat{F}_{\Delta m}^{(n_l+1,2)}(Q,m)$ represents the corrections due to the nonvanishing mass of the heavy quark in the $(n_l+1)$-flavor result and can be cast into the simple form
\begin{align}\label{eq:F_II}
 &\hat{F}_{\Delta m}^{(n_l+1,2)} (Q,m) \nn\\&=  \frac{\big(\alpha_s^{(n_l+1)}\big)^2 C_F T_F}{16\pi^2} \left[f^{(2)}_{m}(\hat{m}) - \left.f^{(2)}_{m}(\hat{m})\right|_{m\rightarrow 0}\right] \, ,
\end{align}
which can be read off explicitly from Eqs.~(\ref{eq:f_QCD}) and~(\ref{eq:F_QCD0}). Due to the fact that $\hat{F}_{\Delta m}^{(n_l+1,2)} (Q,m) \rightarrow 0$ for $\hat{m}\equiv m/Q \to 0$ the massless limit is recovered in the hard function for a vanishing quark mass. The form of Eq.~(\ref{eq:F_II}) is a direct consequence of the fact that the leading IR dependence has to cancel in the SCET matching (in $\MS$). 

The RG evolution from $\mu_H$ to the common scale $\mu$ in the VFNS proceeds in the $(n_l+1)$ flavor scheme as long as the scale is above $\mu_m$ and in the $(n_l)$ flavor scheme if the scale is below $\mu_m$, according to Eq.~\eqref{eq:currentRGE_massless}. Finally, if the RG evolution crosses the flavor matching scale $\mu_m$, one has to account for a threshold correction factor which we call $\mathcal{M}_H^{+}$ if RG evolution crosses from the $(n_l)$ flavor scheme to the $(n_l+1)$ flavor scheme and $\mathcal{M}_H^{-}$ if RG evolution crosses from the $(n_l+1)$ flavor scheme to the $(n_l)$ flavor scheme. They are the inverse of each other since they are just the ratios of the hard function of Eq.~\eqref{eq:hardfunction_at_muH} in the two flavor schemes:
\begin{align}
 \mathcal{M}_H^{+}(Q,m,\mu_m)&=\frac{H^{(n_l+1)}(Q,m,\mu_m)}{H^{(n_l)}(Q,m,\mu_m)}\, ,\label{eq:MH+} \\
  \mathcal{M}_H^{-}(Q,m,\mu_m)&=\frac{H^{(n_l)}(Q,m,\mu_m)}{H^{(n_l+1)}(Q,m,\mu_m)}\, .\label{eq:MH-}
\end{align}
Here the ratios should be expanded with a common choice of either $\alpha_s^{(n_l)}(\mu_m)$ or $\alpha_s^{(n_l+1)}(\mu_m)$. The threshold correction factor for the hard function at fixed $\mathcal{O}(\alpha_s^2)$ reads ($L_m=\ln(m^2/\mu^2)$)
\begin{align}\label{eq:matchingIIb}
 &\left.\mathcal{M}^{-(2)}_H(Q,m,\mu_m)\right|_{\rm FO}
 = \, \frac{\alpha_s^2 C_{\!F} T_F}{16\pi^2}\nn\\&\times\,  \left\{\left[\frac{8}{3}L_m^2+\frac{80}{9}L_m + \frac{224}{27}\right] \ln\bigg(\frac{m^2}{Q^2}\bigg)-\frac{16}{9}L_m^3 -\frac{4}{9}L_m^2\right.  \nn \\
 &\left. \quad+\bigg(\frac{260}{27}+
  \frac{4\pi^2}{3}\bigg)L_m+\frac{875}{27}+\frac{10\pi^2}{9}-\frac{104}{9}\,\zeta_3\right\} \, ,
\end{align}
where $\alpha_s$ can be either written in the $n_l$ or $n_l+1$ scheme and $\mathcal{M}_H^{+(2)}=-\mathcal{M}_H^{-(2)}$. The corrections at fixed $\mathcal{O}(\alpha_s)$ are zero, i.e. $\mathcal{M}_H^{\pm}(Q,m,\mu_m)=1+\mathcal{M}_H^{\pm(2)}+\mathcal{O}(\alpha_s^3)$. Note that the threshold corrections $\mathcal{M}^{\pm}_H$ involve the logarithm $\ln(m^2/Q^2)$. It is directly related to the rapidity divergences arising in the computation of the SCET current with collinear and soft fluctuations tied to the mass shell of the secondary massive quarks, see Ref.~\cite{Gritschacher:2013pha}. In the logarithmic counting $\mathcal{O}(\alpha_s\, \ln(m^2/Q^2))\sim\mathcal{O}(1)$ one therefore has to include also the terms at $\mathcal{O}(\alpha_s^3 \, \ln(m^2/Q^2))$ and $\mathcal{O}(\alpha_s^4 \, \ln^2(m^2/Q^2))$ for a computation of $\mathcal{M}_H^{\pm}$ at precision $\mathcal{O}(\alpha_s^2)$. In Sec.~\ref{sec:PDFmatching} we will show how to obtain these terms via consistency from the PDF threshold correction, for 
which we will also 
demonstrate explicitly the exponentiation property of the rapidity logarithms.

\subsection{VFNS for the Jet Function}

The VFNS for the jet function can be set up in a way analogous to the hard function. The jet function $J^{(n_f)}(s,m,\mu)$ is in the $(n_f)=(n_l+1)$ flavor scheme if $\mu$ is above $\mu_m$ and in the $(n_f)=(n_l)$ flavor scheme if $\mu$ is below $\mu_m$:
\begin{align}
   J^{(n_f)}(s,m,\mu)=\left\{\begin{array}{ll} 
                       J^{(n_l+1)}(s,m,\mu) \hspace{0.5cm} & {\rm for} \,\, \mu > \mu_m \, ,\\
                      \vspace{-0.2cm} \\
                      J^{(n_l)}(s,m,\mu)  \hspace{0.5cm} & {\rm for} \,\, \mu < \mu_m \, .
                      \label{eq:jetfunction_at_mu}
                     \end{array} \right. 
   \end{align}
The common renormalization scale $\mu$ is in general different from the characteristic scale $\mu_J \sim Q\sqrt{1-x}$ of the jet function, so we specify the jet function at the common scale $\mu$ by the jet function at the scale $\mu_J$, which is free of any large logarithmic terms, convoluted with an RG-evolution factor that resums the logarithms between the scales $\mu_J$ and $\mu$. The jet function at the scale $\mu_J$ serves as the initial condition of this RG evolution, and the flavor scheme that is employed for the initial condition depends again on the relation of $\mu_J$ to the flavor matching scale $\mu_m$:
\begin{align}
   J^{(n_f)}(s,m,\mu_J)=\left\{\begin{array}{ll} 
                      J^{(n_l+1)}(s,m,\mu_J) \hspace{0.0cm} & {\rm for} \,\, \mu_J > \mu_m \, , \\
                      \vspace{-0.2cm} \\
                      J^{(n_l)}(s,m,\mu_J)  \hspace{0.0cm} & {\rm for} \,\, \mu_J < \mu_m \, , \label{eq:jetfunction_at_muJ}
                     \end{array} \right. 
   \end{align}
where   
\begin{align}
J^{(n_l)}(s,m,\mu)= & \, J^{(n_l)}(s,\mu) + J^{(n_l,2)}_{m,\rm real}(s,m) \, ,\label{eq:jetmassive0} \\
J^{(n_l+1)}(s,m,\mu)= & \, J^{(n_l+1)}(s,\mu) + J^{(n_l+1,2)}_{\Delta m,\rm dist}(s,m,\mu) \nn\\&+  J^{(n_l+1,2)}_{m,\rm real}(s,m)\,.
\label{eq:jetmassive}
\end{align}
All mass dependent corrections at $\mathcal{O}(\alpha_s^2 C_F T_F)$ can be directly inferred from the results computed in Ref.~\cite{Pietrulewicz:2014qza} due to the fact that the thrust jet function is decomposed out of two hemisphere jet functions each of which are the same as the one in DIS. The terms $J^{(n_l,2)}_{m,\rm real}(s,m)$ and $J^{(n_l+1,2)}_{m,\rm real}(s,m)$ in Eq.~(\ref{eq:jetmassive}) contribute only when the jet invariant
mass is above the threshold $4m^2$ and thus correspond to real production of the massive quarks. They are given by
\begin{align}\label{eq:J_real}
&J^{(n_f,2)}_{m,\rm real}(s,m)=  \, \frac{\big(\alpha_s^{(n_f)}\big)^2 C_F T_F}{16\pi^2}
\frac{1}{s}\, \theta(s-4m^2) \nn\\&\times\,\bigg\{\!-\frac{32}{3}\,\Li_2\bigg(\frac{b-1}{1+b}\bigg) + 
\frac{16}{3}\,\ln{\left(\frac{1-b^2}{4}\right)}\ln\bigg(\frac{1-b}{1+b}\bigg) \nn\\
& -\frac{8}{3}\,\ln^2\bigg(\frac{1-b}{1+b}\bigg)  +\left(\frac{1}{2}\,b^4-\,b^2+\frac{241}{18}\right)\ln\bigg(\frac{1-b}{1+b}\bigg) \nn \\ &- \frac{5}{27}\,b^3+\frac{241}{9}\,b- \frac{8\pi^2}{9} \bigg\}\,,
\end{align}
for both $n_f=n_l$ and $n_f=n_l+1$ with 
\begin{align}
\qquad b = \sqrt{1-\frac{4m^2}{s}} \, .
\label{eq:bdef}
\end{align}
Note that $J^{(n_l,2)}_{m,\rm real}(s,m)$ is zero at the threshold $s=4m^2$ and that it decouples for $m \rightarrow \infty$ automatically due to the threshold $\Theta$-function, as required by the OS prescription. The expression for $J^{(n_l+1,2)}_{\Delta m,\rm dist}(s,m,\mu)$ contains only distributions and corresponds to collinear massive virtual corrections (including soft-bin subtractions) as well as terms related to the subtraction of the massless quark result already contained in $J^{(n_l+1)}(s,\mu)$. Its renormalized expression reads ($\bar s = s/\mu^2$)
\begin{align}
&\mu^2 J^{(n_l+1,2)}_{\Delta m,\rm dist}(s,m,\mu)=  \, \frac{\big(\alpha_s^{(n_l+1)}\big)^2 C_F T_F}{16\pi^2}
\left\{\delta(\bar{s})\left[\frac{8}{9}L_m^3\right.\right. \nn \\ & \left. \left.  +\frac{58}{9}L_m^2+\left(\frac{718}{27}-\frac{8\pi^2}{9}\right)\!L_{m}+\frac{4325}{81}-\frac{58\pi^2}{27} -\frac{32}{3}\zeta_3\right]\right. \nn \\
& \left.+\left[\frac{\theta(\bar{s})}{\bar{s}}\right]_+ \left[-\frac{8}{3}L_{m}^2-
\frac{116}{9}L_{m}-\frac{718}{27}+\frac{8\pi^2}{9}\right] \right. \nn \\ &\left.
+\left[\frac{\theta(\bar{s})\, \ln\,{\bar{s}}}{\bar{s}}\right]_+\left[\frac{16}{3}L_{m}+\frac{116}{9}\right]
-\frac{8}{3}\!\left[\frac{\theta(\bar{s})\,\ln^2{\bar{s}}}{\bar{s}}\right]_+\right\} \, .
\label{eq:J_virt}
\end{align}
The jet function in the $(n_l+1)$ scheme reaches the massless limit, i.e.~$J^{(n_l+1)}(s,m,\mu) \rightarrow J^{(n_l+1)}(s,\mu)$ for $m \rightarrow 0$.

The RG evolution from $\mu_J$ to the common scale $\mu$ in the VFNS proceeds in the $(n_l+1)$ flavor scheme if the scale is above $\mu_m$ and in the $(n_l)$ flavor scheme if the scale is below $\mu_m$, according to Eq.~\eqref{eq:jetRGE_massless}. Finally, if the RG evolution crosses the flavor matching scale $\mu_m$, one has to account for a threshold correction factor which we call $\mathcal{M}_J^{+}$ if RG evolution crosses from the $(n_l)$ flavor scheme to the $(n_l+1)$ flavor scheme and $\mathcal{M}_J^{-}$ if RG evolution crosses from the $(n_l+1)$ flavor scheme to the $(n_l)$ flavor scheme. They  are the inverse of each other since they are just convolutions of the jet function of Eq.~\eqref{eq:jetfunction_at_muJ} in the two flavor schemes:
\begin{align}
 &\mathcal{M}_J^{+}(s,m,\mu_m) \label{eq:MJ+}  \\&=\int \!\df s' \,J^{(n_l+1)}(s-s',m,\mu_m)\!\,\left(J^{(n_l)}\right)^{-1}(s',m,\mu_m) \, ,\nn \\
  &\mathcal{M}_J^{-}(s,m,\mu_m) \label{eq:MJ-}  \\&=\int \!\df s' \,J^{(n_l)}(s-s',m,\mu_m)\!\,\left(J^{(n_l+1)}\right)^{-1}(s',m,\mu_m) \, .\nn
\end{align}
The threshold correction factor for the jet function at fixed $\mathcal{O}(\alpha_s^2)$ reads
\begin{align}
& \left.\mu_m^2 \mathcal{M}^{-(2)}_{J}(s,m,\mu_m)\right|_{\rm FO}=\frac{\alpha_s^2 C_{\!F} T_F}{16\pi^2} 
\left\{\delta(\bar{s})\left[-\frac{8}{9}L_m^3 \right.\right. \nn \\
& -\left.\frac{58}{9}L_m^2-\left(\frac{466}{27}+\frac{4\pi^2}{9}\right)\!L_m -\frac{1531}{54} 
-\frac{10\pi^2}{27}+\frac{80}{9}\,\zeta_3\right]\!  \nn \\
&\left.+\left[\frac{\theta(\bar{s})}{\bar{s}}\right]_+\left[\frac{8}{3}L_m^2+\frac{80}{9}L_m+\frac{224}{27}\right]
\right\} \, ,
\label{eq:matchingIIIb}
\end{align}
where $\alpha_s$ can be either written in the $(n_l)$ or $(n_l+1)$ scheme and $\mathcal{M}_J^{+(2)}=-\mathcal{M}_J^{-(2)}$. The corrections at fixed $\mathcal{O}(\alpha_s)$ are zero, i.e. ${\mathcal{M}_J^{\pm}(s,m,\mu_m)=\delta(s)+\mathcal{M}_J^{\pm(2)}+\mathcal{O}(\alpha_s^3)}$.
Note that $\mathcal{M}^{\pm}_J$ implicitly contain a logarithm $\sim \ln(m^2/s)$ that becomes large for $m\ll\sqrt{s}\sim Q\sqrt{1-x}$ or $m\gg \sqrt{s}$. Its presence becomes more manifest when using the natural scaling variable $\tilde{s}=s/\nu_J^2 \sim \mathcal{O}(1)$ with $\nu_J \sim Q\sqrt{1-x}$ instead of $\bar{s}=s/\mu_m^2$,
\begin{align}
&\left.\nu_J^2 \mathcal{M}^{-(2)}_{J}(s,m,\mu_m,\nu_J)\right|_{\rm FO}=\frac{\alpha_s^2 C_{\!F} T_F}{16\pi^2} 
\left\{\delta(\tilde{s}) \left[-\frac{8}{9}L_m^3 \right.\right. \nn \\
&-\left.\frac{58}{9}L_m^2 -\left(\frac{466}{27}+\frac{4\pi^2}{9}\right)\!L_m -\frac{1531}{54} 
-\frac{10\pi^2}{27}+\frac{80}{9}\,\zeta_3\right] \nn \\
&\left.+
\left(\left[\frac{\theta(\tilde{s})}{\tilde{s}}\right]_++\ln\left(\frac{\nu_J^2}{\mu_m^2}\right)\delta(\tilde{s})\right)\left[\frac{8}{3}L_m^2+\frac{80}{9}L_m+\frac{224}{27}\right]\right\} \, .
\end{align}
 In the logarithmic counting $\mathcal{O}(\alpha_s\, \ln(m^2/s))\sim\mathcal{O}(1)$ one therefore has to include also the terms at $\mathcal{O}(\alpha_s^3 \, \ln(m^2/s))$ and $\mathcal{O}(\alpha_s^4 \, \ln^2(m^2/s))$ for a computation of $\mathcal{M}_J^{\pm}$ at $\mathcal{O}(\alpha_s^2)$. The corresponding results will be discussed in Sec.~\ref{sec:PDFmatching} where we show how to obtain these terms via consistency from the PDF threshold correction.

\subsection{VFNS for the PDF}

Analogous to the case of the hard and the jet function one can set up the VFNS for the PDF. The PDF ${\phi^{(n_f)}(1-z,\mu)}$ is in the $(n_f)=(n_l+1)$ flavor scheme if $\mu$ is above $\mu_m$ and in the $(n_f)=(n_l)$ flavor scheme if $\mu$ is below $\mu_m$:
\begin{align}
   \phi^{(n_f)}(1-z,m,\mu)=\left\{\begin{array}{ll} 
                       \phi^{(n_l+1)}(1-z,m,\mu)  \hspace{0.0cm} & {\rm for} \,\, \mu > \mu_m \, ,\\
                      \vspace{-0.2cm} \\
                    \phi^{(n_l)}(1-z,\mu) \hspace{0.0cm} & {\rm for} \,\, \mu < \mu_m \, .
                     \end{array} \right. 
   \end{align} 
Note that in the $(n_l)$ scheme the dependence on the quark mass vanishes for $m \gg \LQCD$. 
The common renormalization scale $\mu$ is in general different from the characteristic scale $\mu_\phi \sim \LQCD$ of the PDF, so we specify the PDF at the common scale $\mu$ by the PDF at the scale $\mu_\phi$, convoluted with an RG-evolution factor. The PDF at the scale $\mu_\phi$ serves as an initial condition for the RG evolution and since we consider effects of heavy quarks with a mass $m\gg \Lambda_{\text{QCD}}$, the mass scale is always above the scale of the PDF, i.e.~$\mu_m>\mu_\phi$, so that the PDF at $\mu_\phi$ is always in the $(n_l)$ flavor scheme. This is independent of the scaling of $(1-x)$ with respect to $\LQCD/Q$.

The RG evolution from $\mu_\phi$ to the common scale $\mu$ in the VFNS proceeds in the $(n_l+1)$ flavor scheme if the scale is above $\mu_m$ and in the $(n_l)$ flavor scheme if the scale is below $\mu_m$, according to Eq.~\eqref{eq:pdfRGE_massless}. Finally, if the RG evolution crosses the flavor matching scale $\mu_m$, one has to account for a threshold correction factor $\mathcal{M}_\phi^{+}$,
\begin{align}\label{eq:PDFmatchingcoefficient}
 &\mathcal{M}_{\phi}^{+}(1-z,m,\mu_m)\\&=\int \!\df z' \,\phi^{(n_l+1)}(z'-z,m,\mu_m)\!\,\left(\phi^{(n_l)}\right)^{-1}(1-z',m,\mu_m) \, . \nn
 \end{align}
 Since we always assume $\mu_m>\mu_\phi$, only the transition from the $(n_l)$ to the $(n_l+1)$ flavor scheme is relevant for the PDF. The results for $\mathcal{M}^{+}_{\phi}(1-z,m,\mu)$ in the endpoint region can be easily obtained from the well-known PDF threshold factor in the OPE region calculated in Ref.~\cite{Buza:1995ie} and for convenience also given in Eq.~(\ref{eq:Mphiqq2}) of Appendix~\ref{sec:fixed-order} by expanding for $z \rightarrow 1$, which yields
 \begin{align}\label{eq:MPhi2_FO}
  &\left.\mathcal{M}^{+(2)}_{\phi}(1-z,m,\mu_m)\right|_{\rm FO} = \,  \frac{\alpha_s^2 C_F T_F}{(4\pi)^2}\Bigg\{\delta(1-z)\Bigg[2 L_m^2\nn\\&\,+\left(\frac{2}{3}+\frac{8\pi^2}{9}\right)L_m + \frac{73}{18} + \frac{20 \pi^2}{27} - \frac{8}{3}\zeta_3 \Bigg]  \nn \\
  & \,+\left[\frac{\theta(1-z)}{1-z}\right]_+ \left[\frac{8}{3}L_m^2 +\frac{80}{9}L_m + \frac{224}{27}\right]\Bigg\} \, ,
 \end{align}
 where $\alpha_s$ can be either written in the $(n_l)$ or ${(n_l+1)}$ scheme. The corrections at fixed $\mathcal{O}(\alpha_s)$ are zero, i.e. ${\mathcal{M}_{\phi}^{+}(1-z,m,\mu_m)=\delta(1-z)+\mathcal{M}_{\phi}^{+(2)}+\mathcal{O}(\alpha_s^3)}$.
 In Sec.~\ref{sec:PDFmatching} we will also compute this result directly from the definition in Eq.~(\ref{eq:pdf_x1}) for the PDF in the endpoint region. $\mathcal{M}_\phi^+$ contains a large logarithm $\sim\ln{(1-z)}$ that is manifest when rescaling the plus-distribution in terms of the normalized soft momentum variable $\tilde{\ell} \equiv \ell/\nu_\phi$ with $\ell=Q(1-z)$ and $\nu_\phi\sim Q(1-z)$,
  \begin{align}\label{eq:MPhi2_FO2}
    &\left.\frac{\nu_\phi}{Q}\, \mathcal{M}^{+(2)}_{\phi}\bigg(\frac{\ell}{Q},m,Q,\mu_m,\nu_\phi\bigg)\right|_{\rm FO}=  \, \frac{\alpha_s^2 C_F T_F}{(4\pi)^2} \nn\\&\times\,\Bigg\{ \delta(\tilde{\ell})\Bigg[\left(\frac{8}{3}L_m^2 +\frac{80}{9}L_m + \frac{224}{27}\right)\ln\left(\frac{\nu_\phi}{Q}\right)\nn\\&\;\quad+2 L_m^2+\left(\frac{2}{3}+\frac{8\pi^2}{9}\right)L_m  + \frac{73}{18} + \frac{20 \pi^2}{27} - \frac{8}{3}\zeta_3\Bigg]\nn\\&\;\quad+\left[\frac{\theta(\tilde{\ell})}{\tilde{\ell}}\right]_+ \left[\frac{8}{3}L_m^2 +\frac{80}{9}L_m + \frac{224}{27}\right]\Bigg\} \, .
  \end{align}
 The large logarithms $\ln(1-z)$ arise from rapidity divergences in the collinear PDF function $g_{i/P}$ and the soft function $S$ (or csoft function $S_c$) and can not be resummed in the usual RG evolution. In Sec.~\ref{sec:PDFmatching} it will be shown how these logarithms can be resummed using rapidity RG methods as in Refs.~\cite{Chiu:2011qc,Chiu:2012ir} enabling us to evaluate the matching coefficient $\mathcal{M}_\phi^+$ at N$^3$LL order.

\section{Practical Implementation and Consistency Relations}\label{sec:applications}

In this section we specify explicitly the RG properties of the individual matrix elements in the factorization theorem of  Eq.~\eqref{eq:facttheomassive} in the presence of massive quark corrections for all hierarchies between the mass scale and the kinematic scales.
All of the factorization theorems are valid up to power corrections of $\mathcal{O}(1-x)$, independent of the hierarchy between the mass and the kinematic scales, since the change between the OS and $\MS$ renormalization prescriptions does not generate any power corrections involving the mass. Furthermore, we investigate the conceptual implications of using different final renormalization scales which all of the matrix elements are jointly evolved to, which leads to consistency relations among the threshold corrections.

\subsection{Explicit factorization theorems with massive quarks}

 We apply the prescriptions given in Sec.~\ref{sec:massive} and use first for definiteness a common renormalization scale $\mu<\mu_m$. This implies that the matrix elements and couplings are renormalized in the $(n_l)$ scheme at the common $\mu$. If the RG evolution from the natural scale of the matrix element $\mu_i$ (for $i=H,J$) to the final scale $\mu$ crosses the scale $\mu_m$, the scheme is changed leading to the threshold correction $\mathcal{M}^-_i$ and the number of active flavors in the evolution changes from $n_l+1$ to $n_l$. Thus the hard and jet functions can be written as 
\begin{widetext}
  \begin{align}
  H^{(n_l)}(Q,m,\mu)&=\left\{\begin{array}{ll} 
                      H^{(n_l+1)}(Q,m,\mu_H) \, U_H^{(n_l+1)}(Q,\mu_H,\mu_m) \, \mathcal{M}^-_H(Q,m,\mu_m) \, U_H^{(n_l)}(Q,\mu_m,\mu) \hspace{0.0cm} & {\rm for} \,\,  \mu_H > \mu_m \, ,\\
                      \vspace{-0.2cm} \\
                       H^{(n_l)}(Q,m,\mu_H) \, U_H^{(n_l)}(Q,\mu_H,\mu) \hspace{1.0cm} & {\rm for} \,\, \mu_H < \mu_m \, ,
                     \end{array} \right. 
                     \end{align}
                     \begin{align}
   J^{(n_l)}(s,m,\mu)&=\left\{\begin{array}{ll} 
             \int \df s' \int \df s''\int \df s''' \, J^{(n_l)}(s-s',m,\mu_J) \, U_J^{(n_l)}(s'-s'' ,\mu_m,\mu_J) & \\
                      \hspace{2.2 cm} \times \, \mathcal{M}^-_J(s''-s''',m,\mu_m) \, U_J^{(n_l)}(s'''-s'',\mu,\mu_m) \hspace{1.5cm} & {\rm for} \,\, \mu_J > \mu_m \, , \\
                      \vspace{-0.2cm} \\
             \int \df s' \, J^{(n_l)}(s-s',m,\mu_J) \, U_J^{(n_l)}(s',\mu,\mu_J)  \hspace{1.0 cm} & {\rm for} \,\, \mu_J < \mu_m            \, .
                     \end{array} \right. 
  \end{align}
  Note that the PDF is never evolved in the ($n_l+1$) flavor scheme for $\mu<\mu_m$ because we always assume $m \sim \mu_m > \mu_{\phi} \sim \LQCD$, so that Eq.~\eqref{eq:PDFevolution} holds with $n_f=n_l$. The explicit description of the complete factorization theorem with all evolution factors written out thus adopts three different forms depending on the hierarchy between $\mu_m$ on the one hand and $\mu_H$ and $\mu_J$ on the other. For simplicity we set here $\mu=\mu_\phi$, so that the RG factor for the PDF can be dropped. For $\mu_m>\mu_H$ we get
%  \begin{align}
% &F^{\rm I}_1(x,Q,m) = \, \sum\limits_{i=q,\bar{q}} \frac{e_i^2}{2}\,H^{(n_l)}(Q,m,\mu_H)\,  U^{(n_l)}_H(Q,\mu_H,\mu_\phi)\nn \\&\quad\times\, \int \!\df s\!  \int\! \df s'\, J^{(n_l)}(s',m,\mu_J) \, U^{(n_l)}_J(s-s',\mu_\phi,\mu_J)\nn \\
% &\quad \times \phi^{(n_l)}_{i/P}\left(1-x-\frac{s}{Q^2}, \mu_\phi\right)\,.
%\label{eq:FIm}
%  \end{align}
 \begin{align}
 F^{\rm I}_1(x,Q,m) = & \, \sum\limits_{i=q,\bar{q}} \frac{e_i^2}{2}\,H^{(n_l)}(Q,m,\mu_H)\,  U^{(n_l)}_H(Q,\mu_H,\mu_\phi)\, \int \!\df s\!  \int\! \df s'\, J^{(n_l)}(s',m,\mu_J) \, U^{(n_l)}_J(s-s',\mu_\phi,\mu_J) \nn \\
 & \times \phi^{(n_l)}_{i/P}\left(1-x-\frac{s}{Q^2}, \mu_\phi\right)\,.
\label{eq:FIm}
  \end{align}
  This factorization theorem covers in particular the region, where the massive quark decouples and therefore for all renormalizable quantities the OS scheme is used for the secondary massive quark effects. In this regime mass effects in the jet function are power suppressed by $\mathcal{O}\left(\frac{Q^2(1-x)}{m^2}\right)\lesssim\mathcal{O}(1-x)$ and might be dropped due to their small size.\footnote{Since mass effects in the jet sector appear only in the real radiation correction $J^{(n_l)}_{m,\rm real}$ containing a kinematic threshold, they can in practice anyway not contribute.} For $\mu_H>\mu_m>\mu_J$ one gets
%\begin{align}
% &F^{\rm II}_1(x,Q,m)=\,  \sum\limits_{i=q,\bar{q}} \frac{e_i^2}{2}\, H^{(n_l+1)}(Q,m,\mu_H)\nn \\ &\times\, U^{(n_l+1)}_{H}(Q,\mu_H,\mu_m) \, \mathcal{M}^-_{H}(Q,m,\mu_m) \, U^{(n_l)}_{H}(Q,\mu_m,\mu_\phi)
% \nonumber\\
%&\times \int\! \df s \! \int \!\df s'\, J^{(n_l)}(s',m,\mu_J) \,U^{(n_l)}_J(s-s',\mu_\phi,\mu_J)\nn \\&\times \, \phi^{(n_l)}_{i/P}\left(1-x-\frac{s}{Q^2}, \mu_\phi\right)\,.
%\label{eq:FIIm}
%\end{align}
\begin{align}
 F^{\rm II}_1(x,Q,m)= & \,  \sum\limits_{i=q,\bar{q}} \frac{e_i^2}{2}\, H^{(n_l+1)}(Q,m,\mu_H)\, U^{(n_l+1)}_{H}(Q,\mu_H,\mu_m) \, \mathcal{M}^-_{H}(Q,m,\mu_m) \, U^{(n_l)}_{H}(Q,\mu_m,\mu_\phi)
 \nonumber\\
&\times \int\! \df s \! \int \!\df s'\, J^{(n_l)}(s',m,\mu_J) \,U^{(n_l)}_J(s-s',\mu_\phi,\mu_J) \, \phi^{(n_l)}_{i/P}\left(1-x-\frac{s}{Q^2}, \mu_\phi\right)\,.
\label{eq:FIIm}
\end{align}
Here we use both the $\overline{\text{MS}}$ and OS renormalization prescriptions for the secondary massive quark effects in the evolution of the hard function. In particular the $\MS$ scheme allows us to reach the massless limit for the fixed-order hard coefficient $H^{(n_l+1)}(Q,m,\mu_H)$.
Since $\mu_m>\mu_J$ the jet function is still always evolved in the $(n_l)$ scheme. It is easy to see that for $\mu_m=\mu_H$ the two factorization theorems in Eqs.~\eqref{eq:FIm} and \eqref{eq:FIIm} agree due to the matching relation~\eqref{eq:MH-}. So there is a (perturbatively) continuous transition between the two scaling hierarchies described by $F_1^{\rm I}$ and $F_1^{\rm II}$. 

Finally, for $\mu_J>\mu_m>\mu_\phi$ the explicit factorization theorem reads
%\begin{align}
%& F^{\rm III}_1(x,Q,m)= \,  \sum\limits_{i=q,\bar{q}} \frac{e_i^2}{2}\,H^{(n_l+1)}(Q,m,\mu_H)\nn \\&\times\, U^{(n_l+1)}_{H}(Q,\mu_H,\mu_m)  \mathcal{M}^-_{H}(Q,m,\mu_m)\, U^{(n_l)}_{H}(Q,\mu_m,\mu_\phi)
% \nonumber\\
%&\times\int\! \df s  \!\int\! \df s' \!\int \!\df s'' \!\int\! \df s''' \, J^{(n_l+1)}(s''',m,\mu_J) \nn \\&\times\,U^{(n_l+1)}_J(s''-s''',\mu_m,\mu_J) \,\mathcal{M}^-_J(s'-s'',m,\mu_m)\nn \\&\times \,U^{(n_l)}_J(s-s',\mu_\phi,\mu_m) \,  \phi^{(n_l)}_{i/P}\left(1-x-\frac{s}{Q^2}, \mu_\phi\right) \, .
%\label{eq:FIIIm}
%\end{align}
\begin{align}
F^{\rm III}_1(x,Q,m)= & \,  \sum\limits_{i=q,\bar{q}} \frac{e_i^2}{2}\,H^{(n_l+1)}(Q,m,\mu_H)\, U^{(n_l+1)}_{H}(Q,\mu_H,\mu_m)  \mathcal{M}^-_{H}(Q,m,\mu_m)\, U^{(n_l)}_{H}(Q,\mu_m,\mu_\phi)
 \nonumber\\
&\times\int\! \df s  \!\int\! \df s' \!\int \!\df s'' \!\int\! \df s''' \, J^{(n_l+1)}(s''',m,\mu_J)\,U^{(n_l+1)}_J(s''-s''',\mu_m,\mu_J) \,\mathcal{M}^-_J(s'-s'',m,\mu_m)\nn \\&\times \,U^{(n_l)}_J(s-s',\mu_\phi,\mu_m) \,  \phi^{(n_l)}_{i/P}\left(1-x-\frac{s}{Q^2}, \mu_\phi\right) \, .
\label{eq:FIIIm}
\end{align}
\end{widetext}
  Now in addition both renormalization prescriptions are also used for the jet function allowing us to reach the massless limit for the fixed-order structure $J^{(n_l+1)}(s,m,\mu_J)$. In this regime mass corrections in the hard function are power suppressed by $\mathcal{O}\left(\frac{m^2}{Q^2}\right)\lesssim\mathcal{O}(1-x)$ and taking the massless limit might be suitable. Again, it is easy to see that for $\mu_m=\mu_J$ the two factorization theorems in Eqs.~\eqref{eq:FIIm} and \eqref{eq:FIIIm} agree due to the matching relation~\eqref{eq:MJ-}. So there is a (perturbatively) continuous transition between the two scaling hierarchies described by $F_1^{\rm II}$ and $F_1^{\rm III}$.

\subsection{Consistency conditions}

The equivalence of the factorization theorem for different choices of the common renormalization scale $\mu$ concerning physical predictions leads to statements about the intrinsic relations between its components. On the one hand, they imply the well-known consistency conditions between the RG evolution factors $U_H$, $U_J$ and $U_{\phi}$, see Eq.~(\ref{eq:consistency_ML}). On the other hand, 
in the context of the RG evolution crossing a massive quark threshold they also imply a consistency relation between the threshold factors for the hard, jet and parton distribution functions, $\mathcal{M}^\pm_H$, $\mathcal{M}^\pm_J$ and $\mathcal{M}^+_{\phi}$. Apart from providing consistency checks of theoretical calculations, these relations have also computational power, as 
they can be used to calculate properties of independent gauge-invariant field theoretic objects once it has become 
clear that they represent building blocks of a factorization theorem. Hereby, one of the most interesting 
aspects is that the various building blocks can appear in different factorization theorems, and one may gain insights into 
the mass-singularities of apparently unrelated quantities. 

In the previous subsection we have discussed the renormalization of the matrix elements for $\mu<\mu_m$. An equivalent choice would have been $\mu>\mu_m$, where all renormalized quantities are evaluated in the $(n_l+1)$ scheme at the common scale $\mu$. 
Here the RG evolution of the hard and jet functions and the PDF reads
\begin{widetext}
    \begin{align}
   &H^{(n_l+1)}(Q,m,\mu)=\left\{\begin{array}{ll} 
                      H^{(n_l+1)}(Q,m,\mu_H) \, U_H^{(n_l+1)}(Q,\mu_H,\mu)  \hspace{0.5cm} & {\rm for} \,\, \mu_H > \mu_m \, ,\\
                      \vspace{-0.2cm} \\
                      H^{(n_l)}(Q,m,\mu_H) \, U_H^{(n_l)}(Q,\mu_H,\mu_m) \, \mathcal{M}^{+}_H(Q,m,\mu_m) \, U_H^{(n_l+1)}(Q,\mu_m,\mu) \hspace{0.5cm} & {\rm for} \,\, \mu_H < \mu_m \, ,
                     \end{array} \right. 
                     \end{align}
                     \begin{align}
   &J^{(n_l+1)}(s,m,\mu)=\left\{\begin{array}{ll} 
                      \int \df s' J^{(n_l+1)}(s-s',m,\mu_J) \, U_J^{(n_l+1)}(s',\mu,\mu_J) \hspace{1.0 cm} & {\rm for} \,\, \mu_J > \mu_m \, , \\
                      \vspace{-0.2cm} \\
                      \int \df s'   \int \df s'' \int\df s''\int\df s''' J^{(n_l)}(s-s',m,\mu_J) \, U_J^{(n_l)}(s'-s'' ,\mu_m,\mu_J) &  \\
                                            \hspace{2.2cm} \times\, \mathcal{M}^{+}_J(s''-s''',m,\mu_m) \, U_J^{(n_l+1)}(s'''-s'',\mu,\mu_m)  \hspace{1.0cm} & {\rm for} \,\, \mu_J < \mu_m \, ,
                     \end{array} \right. \\
   &\phi^{(n_l+1)}(1-z,m,\mu)= \int \df z' \df z'' \df z''' \phi^{(n_l)}(z'-z,\mu_\phi) \, U_\phi^{(n_l)}(z''-z',\mu_m,\mu_\phi)\nn\\&\hspace{5.0cm}\times\, \mathcal{M}^{+}_\phi(z'''-z'',m,\mu_m)\, U_\phi^{(n_l+1)}(1-z''',\mu,\mu_m) \, .
 \end{align}
 Here the threshold factor $\mathcal{M}^+_{\phi}(1-z,m,\mu_m)$ arises since the RG evolution of the PDF $\phi$ necessarily crosses the massive quark threshold. 
 
 \begin{figure*}
 \centering
  \subfigure{\epsfig{file=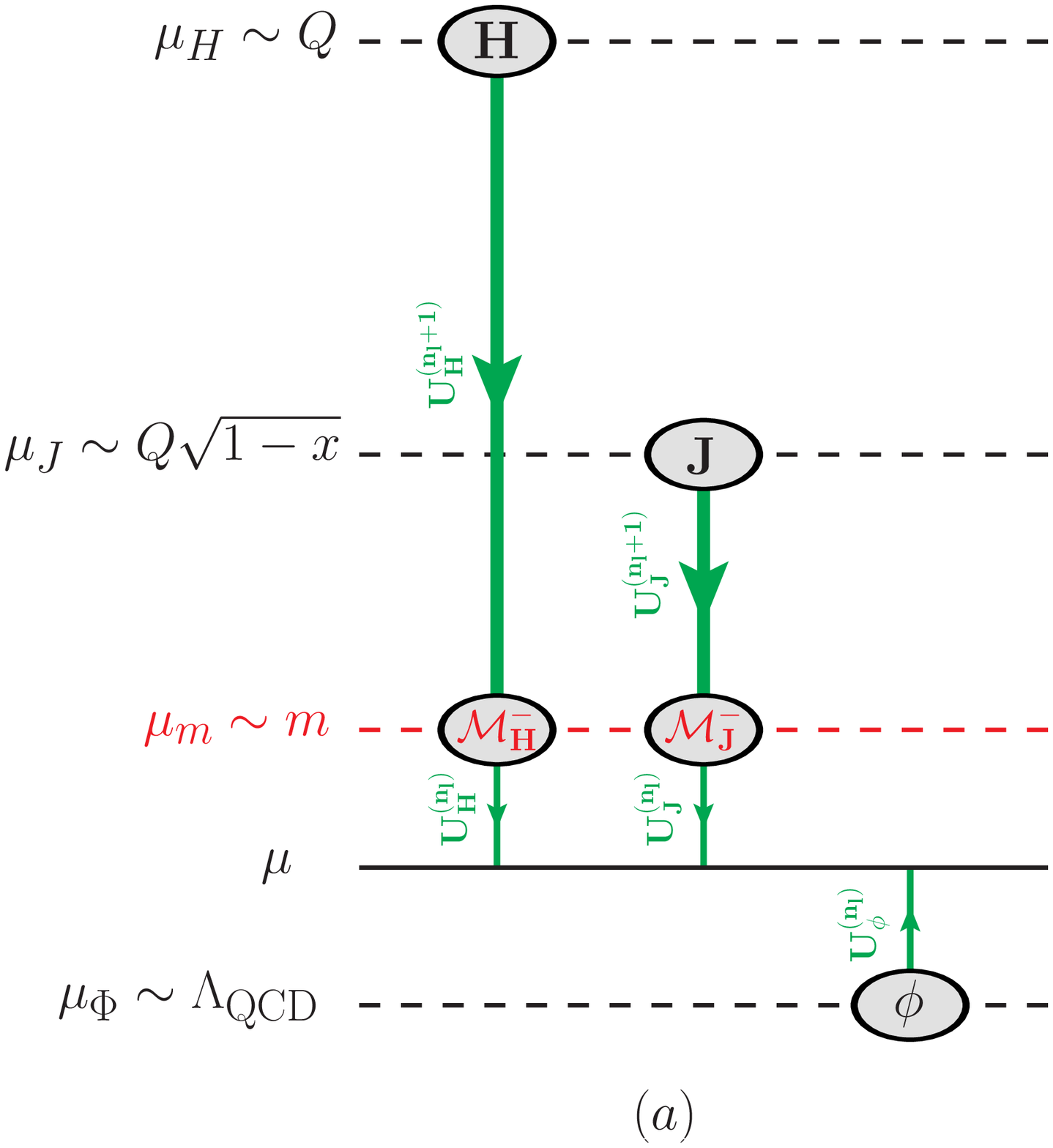,height=0.5\linewidth,clip=}}
  \subfigure{\epsfig{file=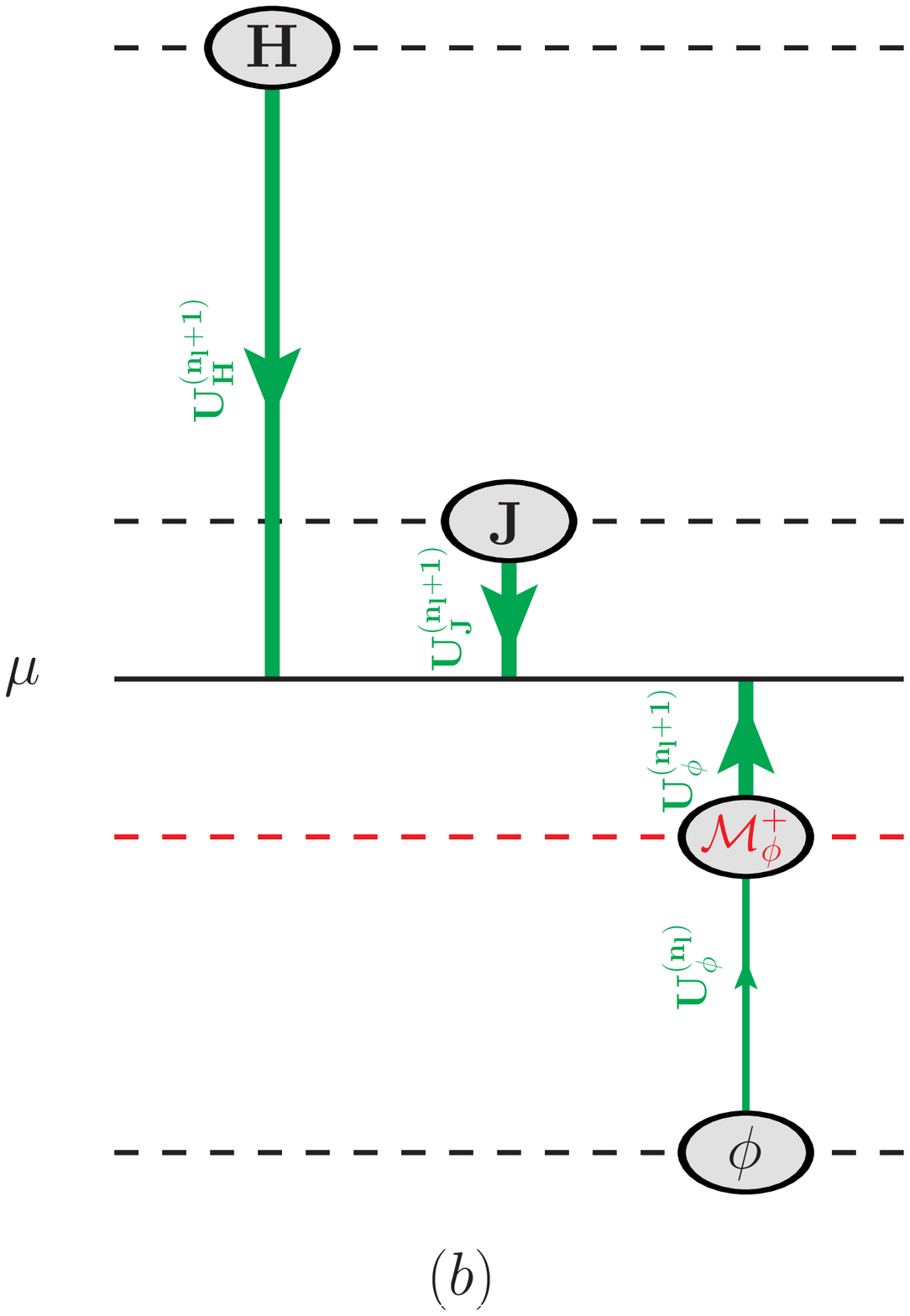,height=0.5\linewidth,clip=}}
   \caption{Illustration of the different RG setups for the hierarchy $\mu_J>\mu_m>\mu_\phi$ leading to the consistency
relations mentioned in the text.
We display the cases where the common renormalization scale $\mu$ satisfies (a) 
$\mu_m>\mu>\mu_\phi$ and (b) $\mu_J>\mu>\mu_m$.} \label{fig:RGconsistency}
\end{figure*}

 In Fig.~\ref{fig:RGconsistency} we show an illustration of the two equivalent choices for $\mu_J>\mu_m$, i.e.~we display the situations where the common renormalization scale $\mu$ lies (a) between the mass and the PDF scales or (b) between the jet and the mass scales.  To discuss the implications let us consider the complete factorization theorem for this specific hierarchy in case (b) with $\mu=\mu_J$ set for simplicity,
%\begin{align}\label{eq:FIIIm2}
%&F^{\rm III}_1(x,Q,m)= \, \sum\limits_{i=q,\bar{q}} \frac{e_i^2}{2}\, H^{(n_l+1)}(Q,m,\mu_H) \\&\times\, U^{(n_l+1)}_{H}(Q,\mu_H,\mu_J)   \, \int\! \df s \, J^{(n_l+1)}(s,m,\mu_J) \nn\\&\times\!\int\! \df z \!\int \!\df z' \, U^{(n_l+1)}_{\phi}(z-x,\mu_J,\mu_m) \, \mathcal{M}^+_{\phi}(z'-z,m,\mu_m)\nn\\&\times \!\int\! \df z''\, U^{(n_l)}_{\phi}(z''-z',\mu_m,\mu_\phi) \,  {\phi}^{(n_l)}_{i/P}\Big(1-z''-\frac{s}{Q^2},\mu_\phi\Big) \, . \nn
%\end{align}
\begin{align}\label{eq:FIIIm2}
F^{\rm III}_1(x,Q,m)= &\, \sum\limits_{i=q,\bar{q}} \frac{e_i^2}{2}\, H^{(n_l+1)}(Q,m,\mu_H) \, U^{(n_l+1)}_{H}(Q,\mu_H,\mu_J)   \, \int\! \df s \, J^{(n_l+1)}(s,m,\mu_J) \int\! \df z  \, U^{(n_l+1)}_{\phi}(z-x,\mu_J,\mu_m) \nn\\&\times\int \!\df z' \!\int\! \df z''\, \mathcal{M}^+_{\phi}(z'-z,m,\mu_m) \, U^{(n_l)}_{\phi}(z''-z',\mu_m,\mu_\phi) \,  {\phi}^{(n_l)}_{i/P}\Big(1-z''-\frac{s}{Q^2},\mu_\phi\Big) \, . 
\end{align}
 \end{widetext}
The equivalence of the factorization theorems in Eqs.~(\ref{eq:FIIIm}) and~(\ref{eq:FIIIm2}) implies, besides the relation between the evolution factors and anomalous dimensions shown in Eq.~(\ref{eq:consistency_ML}) for $n_f=n_l$ and $n_l+1$, also a relation between the threshold correction factors,
\begin{align}\label{eq:consistency_DIS_m}
\mathcal{M}^{+}_{\phi}(1-z,m,\mu)= & \, Q^2  \mathcal{M}^{-}_H(Q,m,\mu) \, \mathcal{M}^{-}_J(Q^2(1-z),m,\mu) \, .
\end{align}
or equivalently
\begin{align}
%&\delta(1-z)=Q^2\mathcal{M}_H^-(Q,m,\mu)\notag \\&\hspace{1cm}\times \int\!\df z' \,\mathcal{M}_J^-(Q^2(1-z'),m,\mu)\mathcal{M}_{\phi}^-(z'-z,m,\mu)\, ,\\
&\delta(1-z)=Q^2\mathcal{M}_H^+(Q,m,\mu) \\&\hspace{0.8cm}\times \int\!\df z' \,\mathcal{M}^+_J(Q^2(1-z'),m,\mu)\, \mathcal{M}_{\phi}^+(z'-z,m,\mu)\, . \nn
\end{align}
These relations imply in particular that the rapidity logarithms (and singularities) that arise in the hard, collinear and soft sectors are intrinsically related to each other. We can explicitly check that the consistency relation is satisfied at $\mathcal{O}(\alpha_s^2)$ in the fixed-order expansion.  Inserting Eqs.~(\ref{eq:matchingIIb}),~(\ref{eq:matchingIIIb}) and (\ref{eq:MPhi2_FO}) confirms Eq.~(\ref{eq:consistency_DIS_m}). We emphasize that for Eq.~(\ref{eq:MPhi2_FO}) to be satisfied for arbitrary masses the coefficients of the rapidity logarithms $\ln(m^2/Q^2)$ in $\mathcal{M}^\pm_H$, $\ln(Q^2(1-x)/m^2)$ in $\mathcal{M}^\pm_J$ and $\ln(1-x)$ in $\mathcal{M}^+_\phi$ need to be equivalent.

\section{Calculation of the PDF threshold correction}\label{sec:PDFmatching}

As already indicated in Eq.~\eqref{eq:PDFmatchingcoefficient}, the PDF threshold correction $\mathcal{M}^+_{\phi}(1-z,m,\mu)$ is given by the ratio of the PDFs in the ($n_l+1$) and ($n_l$) schemes, i.e.
\begin{align}\label{eq:Mphi_cont}
 &\mathcal{M}^+_{\phi} (1-z,m,\mu_m) \nn\\&= \,\int \!\df z' \, \phi^{(n_l+1)}(1-z',m,\mu_m)\! \,\left(\phi^{(n_l)}\right)^{-1}(z'-z,m,\mu_m) \nn \\
 &= \, \int \!\df z' \, Z_{\phi}^{(n_l)}(1-z',m,\mu_m)\! \, \left(Z_{\phi}^{(n_l+1)}\right)^{-1}(z'-z,\mu_m) \, ,
\end{align}
where the second equality arises from the universality of the unrenormalized bare PDF.
Note that in Eq.~(\ref{eq:Mphi_cont}) the calculation of the PDFs can be performed with partonic initial states (i.e.~quarks) since the different renormalization conditions are not affected by the infrared behavior. This can be also seen from the second equality which only involves the renormalization factors. 

Since the PDF in the endpoint region is decomposed out of a soft function $S$ (or a csoft function $S_c$, depending on the scaling of $(1-x)$ with respect to $\LQCD/Q$) and a collinear function $g$, see Eq.~(\ref{eq:pdf_x1}), the analogous relations to Eq.~(\ref{eq:Mphi_cont}) hold also for the corresponding matching coefficients $\mathcal{M}_{S}$ and $\mathcal{M}_{g}$, which are related to $\mathcal{M}^+_{\phi}$ via\footnote{For notational simplicity we use here the soft matching coefficient $\mathcal{M}_S$, which is identical to the csoft matching coefficient $\mathcal{M}_{S_c}$.}
\begin{align}\label{eq:M_dec}
 \mathcal{M}^+_{\phi}(1-z,m,\mu_m)  & =  Q \int \df \ell \, \mathcal{M}_{S}(\ell, m,\mu_m,\nu)   \\ & \times \mathcal{M}_{g}\left(Q(1-z)-\ell,Q,m,\mu_m,\nu\right) \, . \nn
\end{align}
An important technical point is that we encounter rapidity divergences in the calculation of the collinear and soft PDF functions which are not associated to the UV or IR behavior and are not regularized by dimensional regularization. To regulate these divergences we need an additional regulator that breaks boost invariance. Here we display the corresponding results for individual diagrams employing the ``$\eta$-regulator''~\cite{Chiu:2011qc,Chiu:2012ir} for the collinear, soft and csoft Wilson lines, i.e.
\begin{align}\label{eq:regulator}
W_{\bar{n}}=&\sum_{\text{perms}}\mathrm{exp}\left[-\frac{g}{n\cdot \mathcal{P}}\frac{|n\cdot \mathcal{P}|^{-\eta}}{\nu^{-\eta}} \,n \cdot A_{\bar{n}}\right] \,, 
\end{align}
\begin{align}  \label{eq:regulator_S} 
S_{\bar{n}}=&\sum_{\text{perms}}\mathrm{exp}\left[-\frac{g}{\bar{n}\cdot \mathcal{P}}\frac{|2\mathcal{P}_{3}|^{-\eta/2}}{\nu^{-\eta/2}} \,\bar{n}\cdot A_s\right] \,, 
\end{align}
\begin{align}
X_{\bar{n}}=&\sum_{\text{perms}}\mathrm{exp}\left[-\frac{g}{\bar{n}\cdot \mathcal{P}}\frac{|n \cdot \mathcal{P}|^{-\eta/2}}{\nu^{-\eta/2}} \,\bar{n}\cdot A_{cs}\right] \,, \label{eq:regulator_Sc}
\end{align}
and similarly for $S_n$ and $V_{\bar{n}}$, where due to the boost of the csoft modes with respect to the soft modes $|2\mathcal{P}_{3}|$ is replaced by $|n \cdot \mathcal{P}|$. In this context the scale $\nu$ is an auxiliary scale to maintain the dimensions of the regulated integrals which adopts a similar role as the $\mu$ scale in dimensional regularization. In particular, also the strong coupling adopts a $\nu$-scaling proportional to $\eta$.

We follow the method of Refs.~\cite{Chiu:2011qc,Chiu:2012ir} for setting up the rapidity RG evolution. The summation of the rapidity logarithms can be carried out independently after the $\mu$-evolution has been settled which is the approach we are adopting here. We will show that the decomposition in Eq.~(\ref{eq:M_dec}) provides a way to resum rapidity logarithms $\sim \ln(1-x)$ in terms of a RG evolution in $\nu$. A similar factorization in rapidity is used in the hard current matching computation for massive primary quarks of Ref.~\cite{Hoang:2015vua} which also discusses the rapidity RG evolution due to secondary massive quark effects in detail.

 For sufficiently inclusive observables dispersion relations can be used to obtain the results for secondary massive quark radiation (with mass $m$) at $\mathcal{O}(\alpha_s^2 C_F T_F)$ from the results for ``massive gluon'' radiation (with mass $M$) at $\mathcal{O}(\alpha_s)$, which allows us to deal with the technically simpler one-loop computations for the latter instead of performing the two-loop integration directly. The dispersion method has been discussed in detail in Ref.~\cite{Pietrulewicz:2014qza} and we refer to Sec.~IV A therein for the notations and the explicit relations involved. For the following computations we use always Feynman gauge.

\subsection{One-loop results for the PDF soft and collinear functions with a massive gluon}\label{sec:one_loop}

For the computation of the PDF threshold correction we have to consider both the collinear PDF function $g_{q/P}(\ell,\mu)$ defined in Eq.~(\ref{eq:f_qp}) and the csoft or soft functions $S_c(\ell,\mu)$ and $S(\ell,\mu)$ defined in Eqs.~(\ref{eq:csoftfunction}) and~(\ref{eq:S_DIS}), respectively. As already argued in Sec.~\ref{sec:factorization_setup} the csoft and soft function are related by a common boost of the Wilson lines and are therefore in fact equivalent, which we will explicitly show here at the one-loop level. Since we are interested in the matching correction related to different employed renormalization schemes, we can perform the computation with partonic initial states. In this subsection we consider only the massive gluon contributions. A similar calculation has been performed in Ref.~\cite{Fleming:2012kb} within the context of using the gluon mass as an IR regulator.

  \begin{figure}
\centering
 \hspace{-2cm}
  \includegraphics[width=0.7 \linewidth]{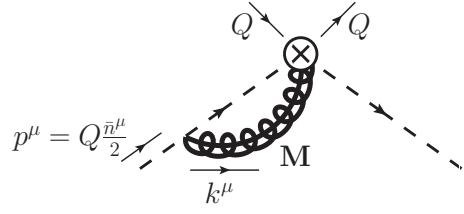}
  \caption{Feynman diagram for the collinear PDF function with a massless quark field in the initial state and a massive gluon at $\mathcal{O}(\alpha_s)$. The symmetric diagram and wave function corrections have to be added. \label{fig:DIS_SCET_M2}} 
   \end{figure}
Let us start with the computation of the partonic collinear contribution $\hat{g}^{(1)}_{q/q}$. Since this is a local matrix element, no real radiation diagrams can contribute.\footnote{To stay in the endpoint region $1-x\ll 1$ the emitted gluon would need to be soft, a contribution that is excluded from the collinear matrix elements by zero-bin subtractions.} Therefore the only contribution (besides the wave-function renormalization) for massive gluon radiation at $\mathcal{O}(\alpha_s)$ is given by the virtual gluon contribution of the diagram in Fig.~\ref{fig:DIS_SCET_M2} and its symmetric configuration, which we denote together by $g_{\bar{n}}$. For convenience, we use a frame where the perpendicular component of the initial onshell quark momentum vanishes, i.e.~$p^\mu=(p^+,p^-,p^{\perp})=(Q,0,0)$. We then obtain
 \begin{align}\label{eq:DIS_M_collinear2}
  &g_{\bar{n}}=  \, 4i g^2 C_F \tilde{\mu}^{2\epsilon} \nu^\eta \delta\left(\ell\right) \int \frac{\df^d k}{(2\pi)^d} \,\frac{Q-k^+}{(k^+)^{1+\eta}} \\&\,\times \,  \frac{1}{[k^+ k^- -\vec{k}_\perp^2 -Q k^- +i\epsilon]} \, \frac{1}{[k^+k^- -\vec{k}_\perp^2-M^2+i\epsilon]}\, . \nn
 \end{align}
After performing the $k^-$- and $k_\perp$-integrations we get 
  \begin{align}
 g_{\bar{n}}=&-\frac{\alpha_s C_F}{\pi} \, \delta(\ell) \,\Gamma\left(2-\frac{d}{2}\right)\left(\frac{\mu^2 e^{\gamma_E}}{M^2}\right)^{2-\frac{d}{2}} \nu^\eta\nn\\&\times\, \int_0^Q \frac{\df k^+}{(k^+)^{1+\eta}} \left(1-\frac{k^+}{Q}\right)^{\frac{d}{2}-1}  \, .
 \end{align}
 This contribution is not regularized by dimensional regularization alone (i.e.~for $\eta=0$) and there is no collinear real radiation contribution to cancel the corresponding rapidity divergence (in contrast to the OPE region $1-x \sim \mathcal{O}(1)$). Expanding for $\eta \rightarrow 0$ gives 
 \begin{align}
  g_{\bar{n}}=&\,\frac{\alpha_s C_F}{\pi} \, \delta(\ell) \,\Gamma\left(2-\frac{d}{2}\right)\left(\frac{\mu^2 e^{\gamma_E}}{M^2}\right)^{2-\frac{d}{2}} \nn\\&\times\,\left\{\frac{1}{\eta}+ \ln\left(\frac{\nu}{Q}\right) +H_{\frac{d}{2}-1}\right\} \, ,
 \end{align}
 where $H_{\alpha}=\psi(1+\alpha)+\gamma_E$ is the Harmonic number.
 The corresponding soft-bin subtractions $g_{\bar{n},0M}$ have to be taken into account, which in general yield some additional corrections. However, for the $\eta$-regulator they vanish. Including the contribution from the wave function renormalization, given by
 \begin{align}
  Z_{\xi}^{(1)} = \frac{\alpha_s C_F}{4\pi}  \,\Gamma\bigg(2-\frac{d}{2}\bigg)\Big(\frac{\mu^2 e^{\gamma_E}}{M^2}\Big)^{2-\frac{d}{2}} \,\frac{2(d-2)}{d} \, ,
 \end{align}
 we obtain in total for the bare partonic collinear function
  \begin{align}\label{eq:f1_qq}
  &\hat{g}^{(\bare,1)}_{q/q}(\ell,M,Q,\mu,\nu) =  \, \left(g_{\bar{n}}-g_{\bar{n},0M}\right)  - Z_{\xi}^{(1)}\, \delta(\ell) \nn \\
 &=  \,  \frac{\alpha_s C_F}{4\pi} \,\delta(\ell) \, \Gamma\left(2-\frac{d}{2}\right) \left(\frac{\mu^2 e^{\gamma_E}}{M^2}\right)^{2-\frac{d}{2}}\nn\\&\quad\quad\times\,\left\{\frac{4}{\eta}+ 4\,\ln\left(\frac{\nu}{Q}\right) +4 H_{\frac{d}{2}-1}- \frac{2(d-2)}{d} \right\} \, .
 \end{align}
 This is in agreement with the results given in Refs.~\cite{Chiu:2012ir,Fleming:2012kb}. 
 
 \begin{figure}
 \centering
 \includegraphics[width=0.99\linewidth]{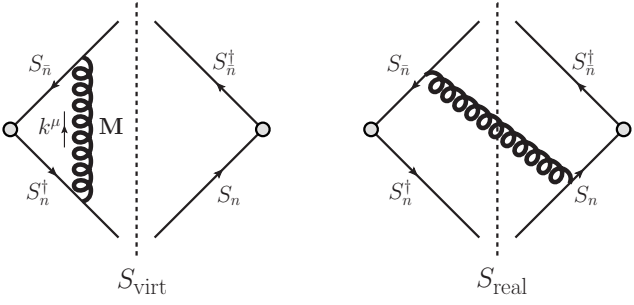}
 \caption{Non-vanishing Feynman diagrams for the computation of the one-loop massive gluon contributions to the soft function $S(\ell,\mu)$. The corresponding symmetric configurations are implied.\label{softfunction_diagrams_DIS}} 
\end{figure}
 
 Let us turn to the computation of the (partonic) soft matrix element $\hat{S}^{(1)}$, where the corresponding diagrams are shown in Fig.~\ref{softfunction_diagrams_DIS}. The virtual diagrams $S_{\text{virt}}$ with the $\eta$-regulator employed for both Wilson lines read (including symmetric configurations) 
  \begin{align}\label{eq:S_virt}
 S_{\text{virt}}=&-4 ig^2 C_F \tilde{\mu}^{2\epsilon}\nu^\eta\,\delta(\ell)\\&\times\,\int \frac{\df^d k}{(2\pi)^d} \, \frac{|k^+-k^-|^{-\eta}}{[k^--i\epsilon] \,[k^+ -i\epsilon]} \,\frac{1}{[k^2-M^2+i\epsilon]}\, .\nn
 \end{align}
 A simple way to evaluate this is to first perform the $k_0$-integration by contours and afterwards the remaining $k_3$- and $k_\perp$-integrations, which yields in agreement with Refs.~\cite{Chiu:2012ir,Fleming:2012kb}
  \begin{align}
  S_{\text{virt}}=&\, -\frac{\alpha_s C_F}{\pi} \, \delta(\ell)\, \Gamma\left(2-\frac{d}{2}\right) \left(\frac{\mu^2 e^{\gamma_E}}{M^2}\right)^{2-\frac{d}{2}}\nn\\&\quad\quad\times\,\left\{\frac{2}{\eta} +2 \,\ln\left(\frac{\nu}{M}\right)+H_{1-\frac{d}{2}}\right\} \, .
 \end{align}
 
 The real radiation diagrams $S_{\text{real}}$ yield
 \begin{align}\label{eq:Sb_1}
 S_{\text{real}}=4 ig^2 C_F &\tilde{\mu}^{2\epsilon}\nu^\eta\int \frac{\df^d k}{(2\pi)^d} \, \frac{|k^+-k^-|^{-\eta}}{[k^--i\epsilon] \,[k^+ -i\epsilon]} \nn\\&\times\, (-2\pi i) \,\delta(k^2-M^2) \, \delta(\ell-k^+) \, .
 \end{align}
 After performing the trivial $k^+$- and $k_{\perp}$-integrations this reads
 \begin{align}\label{eq:Sb_2}
  S_{\text{real}}=& \,\frac{\alpha_s C_F}{\pi} \, \frac{\left(\mu^2 e^{\gamma_E}\right)^{2-\frac{d}{2}}}{\Gamma\left(\frac{d}{2}-1\right)} \, \frac{\theta(\ell)}{\ell}\,\nu^{\eta}\nn\\&\times\, \int_{\frac{M^2}{\ell}}^{\infty} \frac{\df k^{-}}{k^{-}} \,\left(\ell\,k^{-} -M^2\right)^{\frac{d}{2}-2} \,|\ell - k^-|^{-\eta}\, .
 \end{align}
 Finally, expanding in $\eta$ after the $k^-$-integration yields (with $\bar{\ell}\equiv\ell/\nu$)
 \begin{align}\label{eq:Sb_3}
 \nu \, S_{\text{real}}=&\, \frac{\alpha_s C_F}{\pi} \, \Gamma\left(2-\frac{d}{2}\right) \left(\frac{\mu^2 e^{\gamma_E}}{M^2}\right)^{2-\frac{d}{2}}\\&\times\, \left\{\delta(\bar{\ell})\left[\frac{1}{\eta} +2 \,\ln\left(\frac{\nu}{M}\right)+H_{1-\frac{d}{2}}\right]+\left[\frac{\theta(\bar{\ell})}{\bar{\ell}}\right]_+\right\} \nn \, .
 \end{align}
 We note that our computation of $S_{\text{real}}$ differs from Ref.~\cite{Fleming:2012kb} which uses the same regularization methods in several ways: First, our prescription for the Wilson lines in the soft function differs from theirs resulting in a relative sign in Eq.~(\ref{eq:Sb_1}). Second, the result of the phase space integrations in Eq.~(35) of Ref.~\cite{Fleming:2012kb} does not agree with Eq.~(\ref{eq:Sb_3}). Third, we emphasize that in the computation of the soft diagrams we do not encounter any non-vanishing collinear-bin subtractions, in contrast to such a statement given there. However, overall these three deviations cancel each other giving the same result for the total soft real radiation correction in our Eq.~\eqref{eq:Sb_3}.
 
Summing up all contributions the bare soft function reads in terms of $\tilde{\ell}\equiv\ell/\nu_\phi$ with $\nu_\phi \sim Q(1-x)$
\begin{align}\label{eq:S1_qq}
 &\nu_\phi \, S^{(\bare,1)}(\ell,M,\mu,\nu)= \,\nu_\phi \left(S_{\text{virt}}+S_{\text{real}}\right) \nn \\
 &= \, \frac{\alpha_s C_F}{4\pi} \, \Gamma\left(2-\frac{d}{2}\right) \left(\frac{\mu^2 e^{\gamma_E}}{M^2}\right)^{2-\frac{d}{2}}\nn\\&\quad\quad\times\, \left\{\delta(\tilde{\ell})\left[-\frac{4}{\eta} -4 \,\ln\left(\frac{\nu}{\nu_\phi}\right)\right]+ 4 \left[\frac{\theta(\tilde{\ell})}{\tilde{\ell}}\right]_+\right\} 
\end{align}
in agreement with the result in Ref.~\cite{Fleming:2012kb}.
  
Next we will also calculate the csoft function $S_c$ to show that this leads to the same result as for the soft function $S$ above. Here only the rapidity regularization prescription changes, see Eq.~\eqref{eq:regulator_Sc} compared to Eq.~\eqref{eq:regulator_S}. This gives a scaleless contribution for the virtual diagram, such that $S_{c,\text{virt}}=0$. The real radiation diagrams for the csoft function then give
 \begin{align}\label{eq:Screal_1}
  S_{c,\text{real}}=& \,\frac{\alpha_s C_F}{\pi} \, \frac{\left(\mu^2 e^{\gamma_E}\right)^{2-\frac{d}{2}}}{\Gamma\left(\frac{d}{2}-1\right)} \, \frac{\theta(\ell)}{\ell^{1+\eta}}\,\nu^{\eta}\nn\\&\times\, \int_{\frac{M^2}{\ell}}^{\infty} \frac{\df k^{-}}{k^{-}} \,\left(\ell\,k^{-} -M^2\right)^{\frac{d}{2}-2} \nn\\=& \,\frac{\alpha_s C_F}{\pi} \, \Gamma\left(2-\frac{d}{2}\right) \left(\frac{\mu^2 e^{\gamma_E}}{M^2}\right)^{2-\frac{d}{2}} \,\frac{\theta(\ell) \,\nu^\eta}{\ell^{1+\eta}}\, ,
 \end{align}
 which slightly differs from Eq.~(\ref{eq:Sb_2}) concerning the dependence on the $\eta$-regulator.
Expanded in $\eta$  gives exactly the same result as shown in Eq.~\eqref{eq:S1_qq}.
%\begin{align}\label{eq:Screal_2}
%\nu_\phi \, S_{c,\text{real}}
% &= \, \frac{\alpha_s C_F}{\pi} \, \Gamma\left(2-\frac{d}{2}\right) \left(\frac{\mu^2 e^{\gamma_E}}{M^2}\right)^{2-\frac{d}{2}} \\&\quad\quad\times\, \left\{\delta(\tilde{\ell})\left[-\frac{4}{\eta} -4\,\ln\left(\frac{\nu}{\nu_\phi}\right)\right]+4 \left[\frac{\theta(\tilde{\ell})}{\tilde{\ell}}\right]_+\right\} \, , \nn
%\end{align} 
Finally, we remark that at one loop the $\eta$-regulator for the csoft function acts in the same way as the $\alpha$-regulator suggested in Ref.~\cite{Becher:2011dz} applied to the large light-cone component,
\begin{align}
 \frac{\df k^+}{k^+}\to \nu^{\alpha}\,\frac{\df k^+}{(k^+)^{1+\alpha}}\, .
\end{align}
Since the $\alpha$-regulator is boost independent, it gives automatically the same result for the virtual and real radiation diagrams of the soft and csoft function. For the collinear function $g_{\bar{n}}$ both regulators are anyway identical, which thus also implies that the $\eta$-regulator needs to yield the same result for the soft and csoft functions. We will therefore not distinguish between the soft and the csoft functions anymore in the following. 
  
 Expanding Eqs.~(\ref{eq:f1_qq}) and~(\ref{eq:S1_qq}) for $d \rightarrow 4$ gives the unrenormalized corrections\footnote{Here the $\epsilon$-dependence in the expression proportional to $1/\eta$ should be in principle kept unexpanded to avoid terms going like $\epsilon/\eta$ in the $\mu$-anomalous dimension. However, for convenience we show only the terms up 
to $\mathcal{O}(\epsilon^0)$.}  ($L_M=\ln(M^2/\mu^2)$, $\tilde{\ell}=\ell/\nu_{\phi}$)
 \begin{align}\label{eq:g1_qq_bare}
 &\hat{g}^{(\bare,1)}_{q/q}(\ell,M,Q,\mu,\nu)=  \, \frac{\alpha_s C_F}{4\pi} \,\delta(\ell) \nn\\&\times\,\left\{\frac{4}{\eta}\left[\frac{1}{\epsilon}-L_M+\mathcal{O}(\epsilon)\right]+\frac{1}{\epsilon}\left[4\,\ln\left(\frac{\nu}{Q}\right)+3\right]\right. \nn \\
  & \quad\quad\left.-4L_M \, \ln\left(\frac{\nu}{Q}\right)-3L_M +\,\frac{9}{2}-\frac{2\pi^2}{3}\right\} \, , 
  \end{align}
  \begin{align}
  &\nu_\phi \, S^{(\bare,1)} (\ell,M,\mu,\nu) =  \,\frac{\alpha_s C_F}{4\pi} \nn\\&\times\!\left\{\delta(\tilde{\ell})\left(-\frac{1}{\eta}\left[\frac{4}{\epsilon}-4L_M+\mathcal{O}(\epsilon)\right] - \left[\frac{4}{\epsilon}-4L_M\right]\ln\left(\frac{\nu}{\nu_\phi}\right)\right) \right. \nn \\
& \quad\left.+\left[\frac{\theta(\tilde{\ell})}{\tilde{\ell}}\right]_+\left[\frac{4}{\epsilon}-4L_M\right]\right\} \, . \label{eq:S1_qq_bare}
 \end{align}
 We see that $\hat{g}^{(1)}_{q/q}$ and $\hat{S}^{(1)}$ are free of large logarithms for $\nu=\nu_g\sim Q$ and $\nu=\nu_S\sim \nu_\phi \sim Q(1-x)$, respectively. For later reference we also give the resulting $\MS$-type counterterms (subtracting the $1/\epsilon$ and $1/\eta$ divergences)
\begin{align}\label{eq:Zg1}
\nu_\phi \, Z^{(1)}_{g}= & \, \frac{\alpha_s C_F}{4\pi} \,\delta(\tilde{\ell})\left\{\frac{4}{\eta}\left[\frac{1}{\epsilon}-L_M+\mathcal{O}(\epsilon)\right]\right.\nn\\&\left.\quad\,+\frac{1}{\epsilon}\left[4\,\ln\left(\frac{\nu}{Q}\right)+3\right]\right\} \, , \\
\nu_\phi \, Z^{(1)}_{S}=&\, \frac{\alpha_s C_F}{4\pi} \left\{\delta(\tilde{\ell})\left(-\frac{4}{\eta}\left[\frac{1}{\epsilon}-L_M+\mathcal{O}(\epsilon)\right] \right.\right.\nn\\&\left.\left.\,\quad- \frac{4}{\epsilon}\,\ln\left(\frac{\nu}{\nu_\phi}\right)\right)  +\frac{4}{\epsilon}\left[\frac{\theta(\tilde{\ell})}{\tilde{\ell}}\right]_+\right\} \, . \label{eq:ZS1}
\end{align}
%The $\nu$-anomalous dimensions at $\mathcal{O}(\alpha_s)$ can be read off from the counterterm contributions in Eqs.~(\ref{eq:Zg1}) and~(\ref{eq:ZS1}),
%\begin{align}
%\gamma_{g,\nu}^{(1)}= & \,-\nu\frac{\df}{\df\nu}\, Z_{g}^{(1)} = -\frac{\alpha_s C_F}{4\pi} \, 4L_M \,  , \\
%\gamma_{S,\nu}^{(1)}= & \, -\nu\frac{\df}{\df\nu} \, Z_{S}^{(1)} = \frac{\alpha_s C_F}{4\pi} \, 4L_M \, =-\gamma_{g,\nu}^{(1)} \, .
%\end{align}

\subsection{Two-loop results for the PDF soft and collinear functions}

We calculate the secondary massive quark corrections at $\mathcal{O}(\alpha_s^2 C_F T_F)$ for the PDF soft and collinear functions in the $(n_l+1)$ flavor scheme using the (partonic) results for the real and virtual radiation of a massive gluon at $\mathcal{O}(\alpha_s)$ in Eqs.~(\ref{eq:f1_qq}) and~(\ref{eq:S1_qq}). Applying dispersion relations as discussed in Sec.~IV A of Ref.~\cite{Pietrulewicz:2014qza} yields
\begin{align}
 &\hat{g}^{(n_l+1,\bare,2)}_{q/q,T_F}(\ell,m,Q,\Lambda,\mu,\nu)\label{eq:g2bare}\\&= \,\frac{1}{\pi}\int \!\frac{\df M^2}{M^2} \, \hat{g}^{(\bare,1)}_{q/q} (\ell,M,Q,\mu,\nu)\,\mathrm{Im} \left[\Pi(m^2,M^2)\right] \nn\\&\qquad - \left(\Pi(m^2,0)-\frac{\alpha_s^{(n_l+1)} T_F}{3\pi} \,\frac{1}{\epsilon}\right)\! \hat{g}_{q/q}^{(\bare,1)}(\ell,\Lambda,Q,\mu,\nu) \, , \nn 
 \end{align}
 \begin{align}
 &\hat{S}^{(n_l+1,\bare,2)}_{T_F}(\ell,m,\Lambda,\mu,\nu)\label{eq:S2bare}\\&= \,\frac{1}{\pi}\int \!\frac{\df M^2}{M^2} \, \hat{S}^{(\bare,1)}(\ell,M,\mu,\nu)\,\mathrm{Im} \left[\Pi(m^2,M^2)\right] \nn \\
 &\qquad- \left(\Pi(m^2,0)-\frac{\alpha_s^{(n_l+1)} T_F}{3\pi} \,\frac{1}{\epsilon}\right)\! \hat{S}^{(\bare,1)}(\ell,\Lambda,\mu,\nu)\nn \, ,
\end{align}
with the imaginary part of the vacuum polarization function $\Pi(m^2,p^2)$ and its value at zero momentum given by
\begin{align}\label{eq:Im_Pi}
&\mathrm{Im}\!\left[\Pi(m^2,p^2)\right] = \theta(p^2-4m^2) \, g^2 T_F \tilde{\mu}^{2\epsilon}(p^2)^{(d-4)/2} \\&\;\,\times\,\frac{2^{3-2d}\pi^{(3-d)/2}}{\Gamma\Big(\frac{d+1}{2}\Big)} 
\Big(d-2+\frac{4m^2}{p^2}\Big)\!\Big(1-\frac{4m^2}{p^2}\Big)^{(d-3)/2} \! , \nn\\
&\Pi(m^2,0) \,=\,  \frac{\alpha_s T_F}{3\pi} \bigg(\frac{\mu^2 e^{\gamma_E}}{m^2}\bigg)^{2-\frac{d}{2}} 
\Gamma\bigg(2-\frac{d}{2}\bigg) \, .
\label{eq:vacpolzero}
\end{align}
For the scheme change contributions in the respective second terms of Eqs.~\eqref{eq:g2bare} and~\eqref{eq:S2bare} we use a gluon mass $\Lambda \ll m$ as an infrared regulator which allows us to factorize also these corrections with respect to rapidity and to use the results from Sec.~\ref{sec:one_loop}. The total bare results at $\mathcal{O}(\alpha_s^2 C_F T_F)$ for the partonic collinear and soft functions then read
\begin{widetext}
\begin{align}\label{eq:g2}
\nu \, \hat{g}^{(n_l+1,\bare,2)}_{q/q,T_F}= & \, \frac{\big(\alpha_s^{(n_l+1)}\big)^2 C_F T_F}{16\pi^2} \, \delta(\bar{\ell})\left\{\frac{1}{\eta}\left[\frac{8}{3\epsilon^2}-\frac{40}{9\epsilon}-\frac{16}{3}L_m L_\Lambda +\frac{8}{3} L_m^2+\frac{80}{9}L_m+\frac{224}{27} +\mathcal{O}(\epsilon) \right]\right. \nn \\
 & +\frac{1}{\epsilon^2}\left[\frac{8}{3}\,\ln\left(\frac{\nu}{Q}\right)+2\right]+\frac{1}{\epsilon}\left[- \frac{40}{9} \,\ln\left(\frac{\nu}{Q}\right)-\frac{1}{3}-\frac{4\pi^2}{9}\right]+ \left(-\frac{16}{3}L_m L_\Lambda +\frac{8}{3} L_m^2 \right. \nn \\
 & \left.\left.+\,\frac{80}{9}L_m+\frac{224}{27}\right)\ln\left(\frac{\nu}{Q}\right)- 4L_m L_\Lambda+ 2L_m^2+\frac{20}{3}L_m+\frac{73}{18}+\frac{20\pi^2}{27}-\frac{8}{3}\zeta_3\right\}  \, , \\
 \nu \,\hat{S}^{(n_l+1,\bare,2)}_{\phi,T_F}=& \, \frac{\big(\alpha_s^{(n_l+1)}\big)^2 C_F T_F}{16\pi^2} \left\{\frac{1}{\eta} \,\delta(\bar{\ell})\left[-\frac{8}{3\epsilon^2}+\frac{40}{9\epsilon}+\frac{16}{3}L_m L_\Lambda -\frac{8}{3} L_m^2-\frac{80}{9}L_m-\frac{224}{27} +\mathcal{O}(\epsilon) \right]\right. \nn \\
& \left.+\left[\frac{\theta(\bar{\ell})}{\bar{\ell}}\right]_+\left[\frac{8}{3\epsilon^2}-\frac{40}{9\epsilon}-\frac{16}{3}L_m L_\Lambda +\frac{8}{3} L_m^2+\frac{80}{9}L_m+\frac{224}{27}\right]\right\} \, .  \label{eq:S2}
\end{align}
with $\bar{\ell}=\ell/\nu$, $L_m=\ln(m^2/\mu^2)$ and $L_\Lambda=\ln(\Lambda^2/\mu^2)$. Therefore, the nonvanishing two-loop counterterm contributions in $\MS$-renormalization (subtracting also the $1/\eta$-divergences), which are used in the $(n_l+1)$ flavor scheme above the quark mass threshold, read\footnote{We indicate explicitly that only the strong coupling related to the interactions of the gluon to the primary quarks is affected by the rapidity regularization procedure and adopts a $\nu$-dependence. The interactions due to gluon splitting within a single sector do not contain any rapidity divergences and therefore do not need additional regularization. We note that the renormalized strong coupling depends on the scale $\nu$ only due to the dimensional extension of the $k^-$-integration and satisfies $\df\alpha_s(\mu,\nu)/\df\,\ln\,\nu=-\eta\,\alpha_s(\mu,\nu)$ to all orders.}
\begin{align}\label{eq:Zg2}
\nu \, Z_{g,T_F}^{(n_l+1,2)}=& \, \frac{\alpha_s^{(n_l+1)}(\mu,\nu)\,\alpha_s^{(n_l+1)}(\mu)\, C_F T_F}{16\pi^2}\, \delta(\bar{\ell})\left\{\frac{1}{\eta}\left[\frac{8}{3\epsilon^2}-\frac{40}{9\epsilon}-\frac{16}{3}L_m L_\Lambda +\frac{8}{3} L_m^2+\frac{80}{9}L_m +\frac{224}{27} +\mathcal{O}(\epsilon) \right]\right. \nn \\
 & \left.+ \,\frac{1}{\epsilon^2}\left[\frac{8}{3}\,\ln\left(\frac{\nu}{Q}\right)+2\right]+\frac{1}{\epsilon}\left[- \frac{40}{9}\,\ln\left(\frac{\nu}{Q}\right)-\frac{1}{3}-\frac{4\pi^2}{9}\right]\right\} \, , \\
\nu \,Z_{S,T_F}^{(n_l+1,2)}=& \, \frac{\alpha_s^{(n_l+1)}(\mu,\nu)\,\alpha_s^{(n_l+1)}(\mu)\, C_F T_F}{16\pi^2}\left\{\frac{1}{\eta} \,\delta(\bar{\ell})\left[-\frac{8}{3\epsilon^2}+\frac{40}{9\epsilon}+\frac{16}{3}L_m L_\Lambda -\frac{8}{3} L_m^2-\frac{80}{9}L_m -\frac{224}{27} +\mathcal{O}(\epsilon) \right]  \right. \nn \\
&\left. +\left[\frac{\theta(\bar{\ell})}{\bar{\ell}}\right]_+\left[\frac{8}{3\epsilon^2}-\frac{40}{9\epsilon}\right]\right\}\, . \label{eq:ZS2}
\end{align}
\end{widetext}
The sum of the individual counterterm contributions gives the complete PDF counterterm at $\mathcal{O}(\alpha_s^2)$ with respect to one flavor,
\begin{align}
 & Z^{(n_l+1,2)}_{\phi,T_F}\left(\frac{\ell}{Q}=1-z,\mu\right)=  \, Q \left(Z_{g,T_F}^{(n_l+1,2)}+Z_{S,T_F}^{(n_l+1,2)}\right) \nn \\
  &=  \,  \frac{\big(\alpha_s^{(n_l+1)}\big)^2 C_F T_F}{16 \pi^2}\left\{\delta(1-z)\left[\frac{2}{\epsilon^2}-\frac{1}{\epsilon}\left(\frac{1}{3}+\frac{4\pi^2}{9}\right)\right]\right.\nn\\&\left.\qquad+\left[\frac{\theta(1-z)}{1-z}\right]_+\left[\frac{8}{3\epsilon^2}-\frac{40}{9\epsilon}\right]\right\} \, .
\end{align}
This yields, using also the corresponding contributions at $\mathcal{O}(\alpha_s)$ in Eqs.~(\ref{eq:Zg1}) and~(\ref{eq:ZS1}) the correct $\mathcal{O}(\alpha_s^2 C_F T_F)$ contribution to the $\mu$-anomalous dimension for the PDF, 
\begin{align}
  &\gamma^{(n_l+1,2)}_{\phi,T_F}(1-z,\mu) = \frac{\big(\alpha_s^{(n_l+1)}\big)^2 C_F T_F}{16\pi^2}
 \left\{2 \, \Gamma^{(2)}_{T_F} \left[\frac{\theta(1-z)}{1-z}\right]_+  \right.\nn\\&\left.\qquad\,- \left(\frac{4}{3} + \frac{16\pi^2}{9}\right)\! \delta(1-z) \right\} \, . \label{eq:gamma_Phi2}
\end{align}
with $\Gamma^{(2)}_{T_F}=-80/9$ being the $\mathcal{O}(\alpha_s^2 C_F T_F)$ coefficient of the cusp anomalous dimension $\Gamma^{(n_{\!f})}_{{\rm cusp}}$.

%The $\nu$-anomalous dimensions for the pure collinear and the soft PDF contributions in the $\MS$-scheme satisfy
%\begin{align}
%\gamma_{g,\nu,T_F}^{(n_l+1,2)}&= \, -\nu\frac{\df}{\df \nu} \, Z_{g,T_F}^{(n_l+1,2)} \nn\\& = \frac{\big(\alpha_s^{(n_l+1)}\big)^2 C_F T_F}{16\pi^2}\left\{-\frac{16}{3}L_m L_\Lambda +\frac{8}{3} L_m^2\right.\nn\\&\qquad\qquad\left.+\frac{80}{9}L_m+\frac{224}{27}\right\}  \, , \\
%\gamma_{S,\nu,T_F}^{(n_l+1,2)}= &\,-\nu\frac{\df}{\df \nu} \, Z_{S,T_F}^{(n_l+1,2)} =  -\gamma_{g,\nu,T_F}^{(n_l+1,2)} \, .
%\end{align}
%Note that these anomalous dimensions are depending on the infrared regulator $\Lambda$. However, this dependence drops out in the $\nu$-evolution of the massive threshold correction, as we will see below.

Below the mass threshold in the $(n_l)$ flavor scheme the OS subtraction prescription is employed for both the strong coupling and the massive quark contribution to the collinear and soft PDF functions. The OS prescription implies that the secondary massive quark corrections decouple in the limit $m\to \infty$. Since the bare result given in Eq.~\eqref{eq:g2} agrees with its large mass limit it can be easily seen that the $\mathcal{O}(\alpha_s^2C_FT_F)$ massive quark corrections are subtracted away entirely by the OS counterterm. For the PDF threshold corrections at $\mathcal{O}(\alpha_s^2 C_F T_F)$ we therefore obtain
\begin{widetext}
\begin{align}\label{eq:Mg_structure}
\mathcal{M}^{(2)}_{g}(\ell,m,Q,\mu_m,\nu)&=  \left.\int \!\df \ell' \, \hat{g}_{q/q}^{(n_l+1)}(\ell-\ell',m,Q,\Lambda,\mu_m,\nu)\! \,\left(\hat{g}_{q/q}^{(n_l)}\right)^{-1}(\ell',Q,\Lambda,\mu_m,\nu)\right|_{\mathcal{O}(\alpha_s^2)} \nn \\
 &=  \, \hat{g}^{(n_l+1,\bare,2)}_{q/q,T_F}- Z_{g,T_F}^{(n_l+1,2)}- \frac{\alpha_s^{(n_l+1)}T_F}{3\pi}\, L_m\,  \left(\hat{g}_{q/q}^{(\bare,1)}-Z^{(n_l+1,1)}_{g}\right) \, , \\
  \mathcal{M}^{(2)}_{S}(\ell,m,\mu_m,\nu)\nn &=  \left.\int \!\df \ell' \, \hat{S}^{(n_l+1)}(\ell-\ell',m,\Lambda,\mu_m,\nu) \left(\hat{S}^{(n_l)}\right)^{-1}(\ell',\Lambda,\mu_m,\nu)\right|_{\mathcal{O}(\alpha_s^2)} \nn \\
  &=  \, \hat{S}^{(n_l+1,\bare,2)}_{T_F}- Z_{S,T_F}^{(n_l+1,2)}- \frac{\alpha_s^{(n_l+1)}T_F}{3\pi}\, L_m\,  \left(\hat{S}^{(\bare,1)}-Z^{(n_l+1,1)}_{S}\right)\,.\label{eq:MS_structure}
\end{align}
\end{widetext}
 Note that the difference of the scheme for $\alpha_s$ in $\hat{g}_{q/q}^{(n_l)}$, $\hat{S}^{(n_l)}$ and $\hat{g}_{q/q}^{(n_l+1)}$, $\hat{S}^{(n_l+1)}$ affects the terms at $\mathcal{O}(\alpha_s^2 C_F T_F)$ and leads to the third term in the last equality of Eqs.~(\ref{eq:Mg_structure}) and~(\ref{eq:MS_structure}), respectively. Using the two-loop results in Eqs.~(\ref{eq:g2})--(\ref{eq:ZS2}) and the one-loop results in Eqs.~(\ref{eq:g1_qq_bare})--(\ref{eq:ZS1}) with a gluon mass $\Lambda$ as an infrared regulator we obtain
\begin{align}\label{eq:Mf}
&\nu_\phi\, \mathcal{M}^{(2)}_{g}(\ell,m,Q,\mu_m,\nu)=\, \frac{\big(\alpha_s^{(n_l+1)}\big)^2 C_F T_F}{16\pi^2} \, \delta(\tilde{\ell})\nn\\&\;\times\,\left\{\left(\frac{8}{3} L_m^2+\frac{80}{9}L_m+\frac{224}{27}\right)\ln\left(\frac{\nu}{Q}\right) +\,2 L_m^2\right. \nn \\
 &\quad \left.+\left(\frac{2}{3}+\frac{8\pi^2}{9}\right)L_m+\frac{73}{18}+\frac{20\pi^2}{27}-\frac{8}{3}\zeta_3\right\} \, , \\
 &\nu_\phi \, \mathcal{M}^{(2)}_{S}(\ell,m,\mu_m,\nu)=\, \frac{\big(\alpha_s^{(n_l+1)}\big)^2 C_F T_F}{16\pi^2} \, \left(\delta(\tilde{\ell})\,\ln\Bigg(\frac{\nu_\phi}{\nu}\right)\nn\\&\qquad+\left[\frac{\theta(\tilde{\ell})}{\tilde{\ell}}\right]_+ \Bigg)\left[\frac{8}{3} L_m^2 +\frac{80}{9}L_m+\frac{224}{27}\right] \, , \label{eq:MS}
\end{align}
where the dependence on the IR regulator $\Lambda$ has dropped out. Upon summing up $\mathcal{M}^{(2)}_{g}$ and $\mathcal{M}^{(2)}_{S}$ we obtain the total PDF threshold correction already given in Eq.~(\ref{eq:MPhi2_FO}). 

The RGE for the $\nu$-evolution of the threshold corrections reads ($i=g,S$)
 \begin{align}\label{eq:nuevolutionf}
\nu\frac{\df}{\df \nu} \,{\cal M}_{i}(\ell,m,Q,\mu,\nu) & \equiv \gamma_{\mathcal{M}_{i}} \,{\cal M}_{i}(\ell,m,Q,\mu,\nu) \, .
\end{align} 
The $\nu$-anomalous dimensions $\gamma_{\mathcal{M}_{g}}$ and $\gamma_{\mathcal{M}_{S}}$ can be directly read off from Eqs.~(\ref{eq:Mf}) and~(\ref{eq:MS}) or equivalently from the ratio of the counterterms in the ($n_l$) and ($n_l+1$) scheme for $g$ and $S$ in analogy to the last equality in Eq.~\eqref{eq:Mphi_cont}, which gives 
\begin{align}\label{eq:anomalous_dimensions}
 \gamma_{\mathcal{M}_{g}}&=-\gamma_{\mathcal{M}_{S}} 
  \notag\\&=  \frac{\alpha_s^2 C_F T_F}{16\pi^2}\left\{\frac{8}{3} L_m^2+\frac{80}{9}L_m+\frac{224}{27}\right\}+\mathcal{O}(\alpha_s^3)  \, .
\end{align}
The solution of the rapidity RGE in Eq.~(\ref{eq:nuevolutionf}) is a simple exponentiation of the rapidity logarithm, i.e.~setting $\nu=\nu_g$ in $\mathcal{M}_{g}$ and $\nu=\nu_\phi=\nu_S$ in $\mathcal{M}_{S}$ we get
\begin{align}
 &\mathcal{M}_{\phi}(1-z,m,\mu_m)   =  Q \int \df \ell \, \mathcal{M}_{S}(\ell, m,\mu_m,\nu_S)  \nn \\ & \times \mathcal{M}_{g}\left(Q(1-z)-\ell,Q,m,\mu_m,\nu_g\right) \, \left(\frac{\nu_S}{\nu_g}\right)^{\gamma_{\mathcal{M}_{g}}} \, .
\end{align} 
In order to allow for an arbitrary evolution path in $\mu$-$\nu$-space one can generalize this expression to resum the logarithms $\ln(m^2/\mu)$ in the $\nu$-anomalous dimension by  integrating the latter in $\mu$ as discussed in Refs.~\cite{Chiu:2012ir, Hoang:2015vua}.
The variations of the scales $\nu_S$ and $\nu_g$ may be used as an additional input for the perturbative uncertainty estimate. Finally, we remark that by setting $\nu_g=Q$ and $\nu_S=\nu_\phi$ one can obtain the compact all-order expression
\begin{align}\label{eq:M_allorder}
 \mathcal{M}^{+}_\phi(1-z,m,\mu_m)=  \left[\frac{\theta(1-z)}{(1-z)^{1-\gamma_{\mathcal{M}_{g}}}}\right]_+  \gamma_{\mathcal{M}_{g}}\, \mathcal{M}_{\phi,\delta} \, ,
\end{align}
where $\gamma_{\mathcal{M}_{g}}$ is the $\nu$-anomalous dimension with the 2-loop contribution given in Eq.~\eqref{eq:anomalous_dimensions}. $\mathcal{M}_{\phi,\delta}$ denotes the coefficient of the $\delta$-distributions in the PDF threshold correction, i.e.~
\begin{align}
\mathcal{M}_{\phi,\delta} = & \,1+\frac{\alpha_s^2 C_F T_F}{(4\pi)^2}\left[ 2 L_m^2+\left(\frac{2}{3}+\frac{8\pi^2}{9}\right)L_m \right. \nn \\
& +\left.\frac{73}{18}+\frac{20\pi^2}{27}-\frac{8}{3}\zeta_3\right]+ \mathcal{O}(\alpha_s^3)  \, .
\end{align}
The noninteger plus-distribution in Eq.~\eqref{eq:M_allorder} is defined as the analytic continuation of $\theta(1-z)/(1-z)^{1-\gamma_{\mathcal{M}_{g}}}$, see the appendix of Ref.~\cite{Pietrulewicz:2014qza} for details. Expanding Eq.~(\ref{eq:M_allorder}) in $\alpha_s$ allows one to easily to read off the distributive structure of $\mathcal{M}^{+}_\phi$ at any order in the strong coupling in terms of the anomalous dimension $\gamma_{\mathcal{M}_{g}}$.

\subsection{Threshold corrections for N$^3$LL analysis}

For a complete analysis at N$^3$LL we need the terms at $\mathcal{O}(\alpha_s^4\, \ln^2(1-x))$ and $\mathcal{O}(\alpha_s^3\, \ln(1-x))$ both counting as $\mathcal{O}(\alpha_s^2)$ for $\alpha_s \ln(1-x) \sim 1$. The former can be easily obtained from the exponentiation property of the rapidity logarithm. The latter can be read off from the nonsinglet PDF threshold correction in the OPE region that has been recently computed up to $\mathcal{O}(\alpha_s^3)$ in Ref.~\cite{Ablinger:2014vwa}. The corresponding expanded result for $x \rightarrow 1$ (Eq.~(5.60) in Ref.~\cite{Ablinger:2014vwa}) fully agrees with our computation for the $\mu_m$-dependent terms at $\mathcal{O}(\alpha_s^3 \,\ln(1-x))$ (which are obtained from the ratio of the evolution factors in the $(n_l)$ and $(n_l+1)$ flavor schemes), but in addition allows us to extract the relevant $\mu_m$-independent term.

The complete result for the PDF threshold correction in the logarithmic counting $\alpha_s \ln(1-x) \sim 1$ at N$^3$LL reads, ($\tilde{\ell}=\ell/\nu_S,\,\ell\sim\nu_S\sim\nu_\phi\sim Q(1-x),\,\nu_g\sim Q$)
\begin{widetext}
\begin{align}
\frac{\nu_S}{Q}\,\mathcal{M}^+_{\phi}\bigg(\frac{\ell}{Q},m,Q,\mu_m,\nu_g,\nu_S\bigg) =  & \,\, \delta(\tilde{\ell})+\left[\frac{\big(\alpha_s^{(n_l+1)}\big)^2}{(4\pi)^2}\,\delta(\tilde{\ell})\,\ln\left(\frac{\nu_S}{\nu_g}\right) \,\mathcal{M}^{(2)}_{\phi,\ln}(m,\mu_m)\right]_{\mathcal{O}(\alpha_s)} \nn \\
& + \left[\frac{\big(\alpha_s^{(n_l+1)}\big)^2}{(4\pi)^2} \left(\delta(\tilde{\ell}) \,\mathcal{M}_{\phi,1}^{(2)}(Q,m,\mu_m,\nu_\phi,\nu_g)+\left[\frac{\theta(\tilde{\ell})}{\tilde{\ell}}\right]_+ \mathcal{M}^{(2)}_{{\phi},\ln}(m,\mu_m)\right) \right.\nn \\ 
& + \frac{\big(\alpha_s^{(n_l+1)}\big)^3}{(4 \pi)^3} \,\delta(\tilde{\ell})\,\ln\bigg(\frac{\nu_S}{\nu_g}\bigg)\, \mathcal{M}^{(3)}_{{\phi},\ln}(m,\mu_m)  \nn \\
& \left.+\, \frac{\big(\alpha_s^{(n_l+1)}\big)^4}{(4\pi)^4}\,\delta(\tilde{\ell}) \,\ln^2\bigg(\frac{\nu_S}{\nu_g}\bigg)\, \mathcal{M}^{(4)}_{{\phi},\ln^2}(m,\mu_m) \right]_{\mathcal{O}(\alpha_s^2)}+\mathcal{O}(\alpha_s^3) \, ,
\label{eq:matchingPhib}
\end{align}
where the second term in the first line counts formally as $\mathcal{O}(\alpha_s)$ and is therefore already relevant at N$^2$LL. The (universal) functions related to the rapidity logarithms read ($L_m=\ln (m^2/\mu_m^2) $)
\begin{align}\label{eq:M2C}
 \mathcal{M}^{(2)}_{\phi,\ln}(m,\mu_m) = & \, C_F T_F \left\{\frac{8}{3}L_m^2+\frac{80}{9}L_m+\frac{224}{27}\right\} \, , \\
\label{eq:M3C}
 \mathcal{M}^{(3)}_{\phi,\ln}(m,\mu_m) = & \, C_F T_F \left\{L_m^3\!\left[-\frac{176}{27}\,C_{\!A}+\frac{64}{27}\,T_F n_l+\frac{128}{27}\,T_F\right] +L_m^2\!\left[\left(\frac{184}{9}-\frac{16\pi ^2}{9}\right)\! C_{\!A} - 24\,C_{\!F} +\frac{320}{27}\,T_F\right] \right. \nn \\
 &+ L_m\left[\left(\frac{1240}{81}-\frac{160\pi^2}{27}+\frac{224}{3} \zeta_3 \right)\!C_{\!A} +\left(\frac{8}{3}-64\zeta_3\right)\!C_{\!F} +\frac{2176}{81}\,T_F n_l +\frac{1984}{81}\,T_F \right] \nn \\
 & + C_{\!A} \!\left(\frac{35452}{729}- \frac{1648 \pi^2}{243} - 60 \zeta_3  + \frac{176 \pi^4}{135} - \frac{32}{3} B_4 \!\right) + C_{\!F}\! \left(-\frac{2834}{27}+ \frac{1208}{9} \zeta_3 - \frac{16 \pi^4}{15} + \frac{64}{3} B_4 \!\right)  \nn \\
  &  \left.+\,  T_F n_l \!\left(\frac{24064}{729} - \frac{512}{27}\zeta_3\!\right)+  T_F \!\left(-\frac{12064}{729} + \frac{896}{27} \zeta_3\!\right)  \right\} \, , \\
   \mathcal{M}^{(4)}_{\phi,\ln^2}(m,\mu_m) = & \, \frac{\left(\mathcal{M}^{(2)}_{\phi,\ln}(m,\mu_m)\right)^2}{2} = C_F^2 T_F^2 \left\{\frac{32}{9}\,L_m^4+\frac{640}{27}\,L_m^3+\frac{1664}{27}L_m^2+\frac{17920}{243}\,L_m+ \frac{25088}{729}\right\} \, ,
\end{align}
where
 \begin{align}\label{eq:B_4}
 B_4 = \frac{2}{3} \,\ln^4(2) -\frac{2\pi^2}{3} \,\ln^2(2) -\frac{13\pi^4}{180} + 16 \,\Li_4\Big(\frac{1}{2}\Big)\, .
\end{align}
The PDF specific function at $\mathcal{O}(\alpha_s^2)$, which is not multiplied by a rapidity log, reads
\begin{align}
  \mathcal{M}^{(2)}_{\phi,1}(Q,m,\mu_m,\nu_\phi,\nu_g) =   C_F T_F \left\{\mathcal{M}^{(2)}_{\phi,\ln}(m,\mu_m)\,\ln\left(\frac{\nu_g}{Q}\right) +2 L_m^2 +\left(\frac{2}{3}+\frac{8\pi^2}{9}\right)L_m  + \frac{73}{18} + \frac{20 \pi^2}{27} - \frac{8}{3}\zeta_3\right\} \, . 
\end{align}

For completeness we display also the threshold corrections for the hard and jet functions. As discussed in Ref.~\cite{Gritschacher:2013pha} the characteristic rapidity scales for the mass shell fluctuations in $\mathcal{M}^-_{H}$ are $\nu^H_1 \sim Q$ and $\nu^H_2 \equiv \nu_m \sim m$ (with the symmetric $\eta$-regulator), while in $\mathcal{M}^-_{J}$ they are $\nu^J_1 \sim Q$ and $\nu^J_2 \sim m^2/(Q(1-x))$. To account for correlations between these scales we set $\nu^H_1=\nu^J_1=\nu_g$ and $\nu^J_2 = \nu_m^2/\nu_S$. With these choices the results at N$^3$LL have the form
\begin{align}
\mathcal{M}^-_{H}\left(Q,m,\mu_m,\nu_g,\nu_m\right) =   &\,\, 1+\left[\frac{\big(\alpha_s^{(n_l+1)}\big)^2}{(4\pi)^2}\,\ln\left(\frac{\nu_m^2}{\nu_g^2}\right) \,\mathcal{M}^{(2)}_{H,\ln}(m,\mu_m)\right]_{\mathcal{O}(\alpha_s)} \nn \\
& + \left[\frac{\big(\alpha_s^{(n_l+1)}\big)^2}{(4\pi)^2} \mathcal{M}_{H,1}^{(2)}(Q,m,\mu_m,\nu_g,\nu_m) + \frac{\big(\alpha_s^{(n_l+1)}\big)^3}{(4 \pi)^3} \,\ln\left(\frac{\nu_m^2}{\nu_g^2}\right) \mathcal{M}^{(3)}_{H,\ln}(m,\mu_m)\right.  \nn \\
 & \left.\qquad +\frac{\big(\alpha_s^{(n_l+1)}\big)^4}{(4\pi)^4} \,\ln^2\left(\frac{\nu_m^2}{\nu_g^2}\right) \mathcal{M}^{(4)}_{H,\ln^2}(m,\mu_m) \right]_{\mathcal{O}(\alpha_s^2)}+\mathcal{O}(\alpha_s^3) \, ,
\label{eq:matchingII}
\end{align}
 and ($\tilde{s}=s/(\nu_g\nu_S)$)
\begin{align}
\nu_g\nu_S\, &\mathcal{M}^-_J\left(s,m,\mu_m,\nu_g,\frac{\nu^2_m}{\nu_S}\right) =  \,\, \delta(\tilde{s})+\left[\frac{\big(\alpha_s^{(n_l+1)}\big)^2}{(4\pi)^2}\,\delta(\tilde{s})\,\ln\left(\frac{\nu_g \nu_S}{\nu_m^2}\right) \mathcal{M}^{(2)}_{J,\ln}(m,\mu_m)\right]_{\mathcal{O}(\alpha_s)}  \nn \\
& \quad+ \left[\frac{\big(\alpha_s^{(n_l+1)}\big)^2}{(4\pi)^2} \Bigg(\delta(\tilde{s}) \,\mathcal{M}_{J,1}^{(2)}(m,\mu_m,\nu_m) +\left[\frac{\theta(\tilde{s})}{\tilde{s}}\right]_+ \mathcal{M}^{(2)}_{J,\ln}(m,\mu_m)\Bigg)+ \frac{\big(\alpha_s^{(n_l+1)}\big)^3}{(4 \pi)^3}\,\delta(\tilde{s})\, \,\ln\left(\frac{\nu_g \nu_S}{\nu_m^2}\right) \mathcal{M}^{(3)}_{J,\ln}(m,\mu_m) \right.\nn \\ 
& \quad\qquad+ \left. \frac{\big(\alpha_s^{(n_l+1)}\big)^4}{(4\pi)^4}\,\delta(\tilde{s}) \,\ln^2\left(\frac{\nu_g \nu_S}{\nu_m^2}\right) \mathcal{M}^{(4)}_{J,\ln^2}(m,\mu_m) \right]_{\mathcal{O}(\alpha_s^2)}+\mathcal{O}(\alpha_s^3) \, . 
\label{eq:matchingIII}
\end{align}\end{widetext}
As stated at the end of Sec.~\ref{sec:applications} the consistency relation~(\ref{eq:consistency_DIS_m}) implies that the coefficients of the rapidity logarithms, i.e.~the $\nu$-anomalous dimensions, are the same for all threshold corrections, i.e.~
\begin{align}
 \mathcal{M}^{(2)}_{H,\ln}(m,\mu_m) = &\,  \mathcal{M}^{(2)}_{J,\ln}(m,\mu_m) =   \mathcal{M}^{(2)}_{\phi,\ln}(m,\mu_m) \, , \\
 \mathcal{M}^{(3)}_{H,\ln}(m,\mu_m) = &\,  \mathcal{M}^{(3)}_{J,\ln}(m,\mu_m) =   \mathcal{M}^{(3)}_{\phi,\ln}(m,\mu_m) \, , \\
 \mathcal{M}^{(4)}_{H,\ln^2}(m,\mu_m)= &\, \mathcal{M}^{(4)}_{J,\ln^2}(m,\mu_m) =   \mathcal{M}^{(4)}_{\phi,\ln^2}(m,\mu_m) \, ,
\end{align}
which has been already used implicitly in Ref.~\cite{Pietrulewicz:2014qza}. Finally, the remaining function-specific $\mathcal{O}(\alpha_s^2)$ corrections read
\begin{align}
 &\mathcal{M}_{H,1}^{(2)}(Q,m,\mu_m,\nu_g,\nu_m) = \, C_F T_F \left\{-\frac{16}{9}L_m^3-\frac{4}{9}L_m^2\right.\nn\\&\left.+ \left(\frac{260}{27}+\frac{4\pi^2}{3}\right)\!L_m + \frac{875}{27}+\frac{10\pi^2}{9}-\frac{104}{9}\zeta_3 \right. \nn \\
&  \left.+\, \mathcal{M}^{(2)}_{H,\ln}(m,\mu_m)\left[ \ln\bigg(\frac{\nu_g^2}{Q^2}\bigg)-\ln\bigg(\frac{\nu_m^2}{m^2}\bigg)\right]\right\} \, , \\
 &\mathcal{M}^{(2)}_{J,1}(m,\mu_m,\nu_m) = \, C_F T_F \left\{-\frac{8}{9}L_m^3 - \frac{58}{9}L_m^2\right.\nn\\&\quad-\left(\frac{466}{27}+\frac{4\pi^2}{9}\right)\!L_m -\frac{1531}{54}-\frac{10\pi^2}{27}+\frac{80}{9}\zeta_3 \nn \\
 &\left. \quad+ \, \mathcal{M}^{(2)}_{J,\ln}(m,\mu_m) \,\ln\bigg(\frac{\nu_m^2}{\mu_m^2}\bigg)\right\} \, .
\end{align}
Since hard and jet functions and PDFs are building blocks of factorization theorems for many different processes at hadron-hadron collisions, the results for the threshold corrections can be directly applied there as well.

\section{Conclusions}\label{sec:conclusions}

In this work we have discussed how to set up a VFNS for multiscale processes at hadronic collisions, where we have taken inclusive DIS in the endpoint region $x \to 1$ as a specific example. In this limit massive quarks do not (predominantly) participate directly in the hard interaction with the virtual photon and therefore mainly arise as secondary radiation giving corrections starting at $\mathcal{O}(\alpha_s^2)$ in the fixed-order expansion. Starting from the massless factorization theorem we have shown how to systematically incorporate the secondary massive quark effects by using two kinds of renormalization conditions for the massive quark corrections in the gauge invariant components. The use of the $\MS$ renormalization prescription in the small mass region and of the on-shell (low momentum subtraction) renormalization prescription in the large mass region imply automatically that all large logarithms are resummed and that the respective correct limiting behavior is achieved in the massless and the 
decoupling regions. The 
difference between these two schemes manifests itself in additional threshold matching corrections at the mass scale. We have discussed some universal features of these threshold corrections, which exhibit intrinsic relations among each other due to the consistency of RG running. Here we have 
also computed explicitly the PDF threshold correction for $x \to 1$ at $\mathcal{O}(\alpha_s^2)$ and showed how to resum a large remaining logarithm therein that is related to the separation of mass shell fluctuations along rapidity and displayed final expressions for a N$^3$LL analysis. From a practical point of view our VFNS in the endpoint region of DIS can be combined with a VFNS in the OPE region $1-x \sim \mathcal{O}(1)$ by adding the known associated nonsingular corrections related to the difference between the full perturbative QCD result and the fixed-order expressions for the components of the SCET factorization theorem for $x \to 1$. This may have an effect also for moderate values of $x$ due to dynamical threshold enhancement (see e.g.~Ref.\cite{Becher:2007ty}), an effect which reinforces perturbative corrections close to the partonic threshold due to the steep fall-off of the PDFs for momentum fractions close to one.

While we have concentrated in this work on DIS, the concept of how to theoretically treat the effects of secondary massive quarks within factorization is applicable for more general processes including hadron-hadron collisions. In particular, the massive quark threshold corrections relevant for resummation of logarithms at N$^3$LL order determined for the massive components in the factorization theorem (hard function, jet function, PDF) are universal and can be employed in factorization theorems for other processes where these components appear.

\begin{acknowledgments}
 We would like to thank Ilaria Jemos for collaboration at an early stage of this
work. The work of P.P. was supported by the German Science Foundation (DFG) under the Collaborative Research Center (SFB) 676 Particles, Strings and the Early Universe. D.S. is supported by the FWF Austrian Science Fund under the Doctoral Program No. W1252-N27 Particles and Interactions.
\end{acknowledgments}

\appendix

\section{Secondary massive quark corrections in the OPE region}\label{sec:fixed-order}

We display explicit results for the perturbative corrections due to secondary massive quarks at $\mathcal{O}(\alpha_s^2)$ in the classical OPE region, where $1-x \sim \mathcal{O}(1)$. Here the familiar factorization theorem for $n_f$ massless quarks reads
\begin{align}\label{eq:fact_theorem_classical}
F_{1}(x,Q) = \sum\limits_{i=q} \frac{e_i^2}{2} \sum\limits_{j=q,g} \int_x^1 \frac{d \xi}{\xi} \,H^{(n_f)}_{ij}\left(\frac{x}{\xi},\mu\right) f^{(n_f)}_{j/P}(\xi,\mu) \, ,
\end{align}
where the index $q$ includes both quarks and antiquarks. The factorization theorem for $F_{2}(x,Q)$ is analogous. Note, however, that the Callan-Gross relation $F_2=2x F_1$ does not hold in the OPE region beyond tree level.
In the Breit frame the PDFs $f^{(n_f)}_{j/P}$ are forward matrix elements of SCET operators decomposed out of collinear fields~\cite{Bauer:2002nz,Stewart:2010qs}. For definiteness we set the final renormalization scale to be $\mu=\mu_H$. In the massive quark case one obtains
\begin{align}\label{eq:fact_theorem_M1}
 &F^{\rm I}_1(x,Q,m) = \sum\limits_{i=q,Q} \frac{e_i^2}{2} \sum\limits_{j,k=q,g} \int \frac{\df \xi}{\xi} \int \frac{\df \xi'}{\xi'} \\&\times H^{(n_l)}_{ij}\left(\frac{x}{\xi},Q,m,\mu_H\right) U^{(n_l)}_{f,jk}\left(\frac{\xi}{\xi'},\mu_H,\mu_f\right) f^{(n_l)}_{k/P}(\xi',\mu_f) \, . \nn 
\end{align}
for $\mu_m \gtrsim \mu_H$ and
\begin{align}\label{eq:fact_theorem_M2}
 &F^{\rm II}_1(x,Q,m) \\&=  \, \sum\limits_{i=q,Q} \frac{e_i^2}{2} \sum\limits_{j,k=q,Q,g} \,\sum\limits_{l,m=q,g} \int \frac{\df \xi}{\xi} \int \frac{\df \xi'}{\xi'}\int \frac{\df \xi''}{\xi''} \int \frac{\df \xi'''}{\xi'''} \nn\\&\times H^{(n_l+1)}_{ij}\left(\frac{x}{\xi},Q,m,\mu_H\right) U^{(n_l+1)}_{f,jk}\left(\frac{\xi}{\xi'},\mu_H,\mu_m\right)\nn\\
 & \times  \mathcal{M}_{f,kl}\left(\frac{\xi'}{\xi''},m,\mu_m\right) U^{(n_l)}_{f,lm}\left(\frac{\xi''}{\xi'''},\mu_m,\mu_f \right)  f^{(n_l)}_{m/P}(\xi''',\mu_f) \, . \nn
\end{align}
for $\mu_m \lesssim \mu_H$. The secondary massive quark corrections to the hard functions $H^{(n_l)}_{qq}$ and $H^{(n_l+1)}_{qq}$ at $\mathcal{O}(\alpha_s^2 C_F T_F)$ can be written as 
\begin{align}
 H^{(n_l)}_{qq}(z,Q,m,\mu)= & \, H^{(n_l)}_{qq}(z,Q,\mu) + \hat{F}^{(n_l,2)}_{1,m}(z,Q,m) \, , \\
 H^{(n_l+1)}_{qq}(z,Q,m,\mu)= & \, H^{(n_l+1)}_{qq}(z,Q,\mu) + \hat{F}^{(n_l+1,2)}_{1,\Delta m}(z,Q,m) \, .
\end{align}
The full QCD result at ${\cal O}(\alpha_s^2 C_F T_F)$ is both IR- and UV-finite and can be decomposed into a purely virtual correction and a real radiation correction with the kinematic threshold $z=1/(1+4\hat{m}^2)$,
 \begin{align}\label{eq:Fnl_tot}
 &\hat{F}^{(n_l,2)}_{1,m}(z,Q,m) = \,2  \hat{F}^{(n_l,2)}_{m}(Q,m) \,\delta(1-z) \nn\\&\qquad+ \theta(z) \theta(1-z-4 \hat{m}^2 z)\, \hat{F}^{(n_l,2)}_{1,m,\theta}(z,Q,m) \, ,
 \end{align}
  where $\hat{m}=m/Q$.
The partonic QCD current form factor $\hat{F}^{(n_l,2)}_{m}(Q,m)$ is given in Eq.~(\ref{eq:F_QCD}). The real radiation function $\hat{F}^{(n_l,2)}_{m,\theta}(z,Q,m)$ was first computed in Ref.~\cite{Buza:1995ie} (and also checked by us) and reads
\begin{widetext}
\begin{align}\label{eq:f_QCD_theta}
\hat{F}^{(n_l,2)}_{m,\theta}(z,Q,m)  = & \,\, \frac{\big(\alpha_s^{(n_l)}\big)^2 C_{\!F} T_F}{16\pi^2}\, \frac{1}{1-z}\left\{\frac{8}{3} \left[1+z^2-12\hat{m}^4 z^2(1-3z+3z^2)\right] \left[\Li_2\left(\frac{r_z-w_z}{r_z+1}\right)+\Li_2\left(\frac{r_z+w_z}{r_z-1}\right)\right.\right. \nn\\
& -\left.\Li_2\left(\frac{r_z-w_z}{r_z-1}\right)- \Li_2\left(\frac{r_z+w_z}{r_z+1}\right)+\ln\left(\frac{1+r_z}{1-r_z}\right)\ln\left(\frac{r_z+w_z}{r_z-w_z}\right)\right] \nn\\
& + \frac{8}{9} \, r_z \left[-8-11 z^2+ 2\hat{m}^2 z (13-18z+28z^2)\right] \ln\left(\frac{r_z+w_z}{r_z-w_z}\right) \nn \\
&+  \frac{4}{3(1-z)^2}\left[1-3 z^2 +2z^3 +6 \hat{m}^4 z^2 (1-2 z) (7-12z+6z^2)\right]\ln\left(\frac{1+w_z}{1-w_z}\right) \nn \\
& \left.+\,\frac{2w_z}{27(1-z)} \left[151-265z+436 z^2-322 z^3 - 2 \hat{m}^2 z(491 -1530 z+2030z^2 -996z^3)\right]\right\} \, .
\end{align}
\end{widetext}
Here we have used the abbreviations
\begin{align}
r_z=\sqrt{1-4 \hat{m}^2 z} \, , \quad w_z=\sqrt{1-\frac{4 \hat{m}^2 z}{1-z}} \, .      
\end{align}
We remark that the QCD corrections decouple in the heavy quark limit using the $n_l$ scheme for $\alpha_s$, i.e.~$ \hat{F}^{(n_l,2)}_{1,m}(z,Q,m) \rightarrow 0$ for $\hat{m} \to \infty$. In the small mass limit $\hat{m} \rightarrow 0$, on the other hand, we obtain
\begin{widetext}
\begin{align} \label{eq:f_QCD0}
\left.\hat{F}^{(n_l,2)}_{1,m}(z,Q,m) \right|_{m\rightarrow 0} = & \, \frac{\big(\alpha_s^{(n_l)}\big)^2 C_{\!F} T_F}{16\pi^2} \, \theta(z) \, \theta(1-z) \left\{\delta(1-z)\left[2\,\ln^2{(\hat{m}^2)} +\bigg(\frac{38}{3}+\frac{16\pi^2}{9}\bigg)\ln{(\hat{m}^2)}+\frac{265}{9} + \frac{134\pi^2}{27}\right] \right. \nn \\
&+\left[\frac{1}{1-z}\right]_+\left[\frac{8}{3}\,\ln^2(\hat{m}^2)+\frac{116}{9}\,\ln(\hat{m}^2)+\frac{718}{27}-\frac{8\pi^2}{9}\right]+\left[\frac{\ln(1-z)}{1-z}\right]_+\left[-\frac{16}{3}\, \ln(\hat{m}^2)-\frac{116}{9}\right] \nn \\
&+ \frac{8}{3} \left[\frac{\ln^2(1-z)}{1-z}\right]_+ -\frac{8(1+z^2)}{3(1-z)}\,\Li_2(1-z)+ \frac{4 (1 + z^2)}{1-z} \,\ln^2(z) -\frac{4}{3}\left(1 + z\right) \ln^2(1-z)\nn \\
& - \frac{16(1+z^2)}{3(1-z)} \, \ln(z) \, \ln(1-z) +\frac{4}{9 (1-z)}\ln(z) \left[12 \left(1+z^2\right) \ln(\hat{m}^2)+ 29-6z+44 z^2 \right] \nn \\
& +\frac{8}{9} \,\ln(1-z) \left[3 \left(1 + z\right) \ln(\hat{m}^2)+8 + 11 z\right] - \frac{4}{3}\left(1+z\right) \ln^2(\hat{m}^2) -\frac{8}{9}\left(8+11z\right)\ln(\hat{m}^2)\nn \\
& \left.-\,\frac{416}{27} - \frac{644}{27} z + \frac{4\pi^2}{9} (1 + z) \right\}\, .
\end{align}
\end{widetext}
The function $\hat{F}^{(n_l+1,2)}_{1,\Delta m}$ represents the quark mass correction to the massless quark result in the ($n_l+1$) flavor scheme and\footnote{Here the superscript $n_l+1$ indicates only that the ($n_l+1$) scheme for $\alpha_s$ is used in the expressions (\ref{eq:Fnl_tot}) and (\ref{eq:f_QCD0}).}
\begin{align}\label{eq:C2_MS}
 &\hat{F}^{(n_l+1,2)}_{1,\Delta m}(z,Q,m)  \\&\qquad=   \,\hat{F}^{(n_l+1,2)}_{1,m}(z,Q,m) -\left.\hat{F}^{(n_l+1,2)}_{1,m}(z,Q,m) \right|_{m\rightarrow 0} \,, \nn
\end{align}
which vanishes in the massless limit.

The PDF threshold correction $\mathcal{M}^{(2)}_{f,qq}$ is given by~\cite{Buza:1995ie} 
\begin{align}\label{eq:Mphiqq2}
 &\mathcal{M}^{(2)}_{f,qq}(z,m,\mu_m) =  \,  \frac{\alpha_s^2 C_F T_F}{(4\pi)^2} \,\theta(z) \,\theta(1-z)\nn\\&\times\left\{\delta(1-z)\left[2 L_m^2 +\left(\frac{2}{3}+\frac{8\pi^2}{9}\right)L_m +\frac{73}{18} + \frac{20 \pi^2}{27} - \frac{8}{3}\zeta_3 \right] \right. \nn \\
 &  \quad\left.+ \left[\frac{1}{1-z}\right]_+ \left[\frac{8}{3}L_m^2 +\frac{80}{9}L_m + \frac{224}{27}\right] - \frac{4}{3}L_m^2 (1+z) \right.\nn\\&\left.\quad+L_m\left[\frac{8}{9}-\frac{88}{9}z+\frac{8(1+z^2)}{3(1-z)}\,\ln(z)\right]+\frac{2(1+z^2)}{3(1-z)}\,\ln^2(z) \right.\nn \\
 &  \left.\quad+ \frac{\ln(z)}{1-z}\left[\frac{44}{9}-\frac{16}{3}z+\frac{44}{9}z^2\right] +\frac{44}{27}- \frac{268}{27} z \right\} \, ,
\end{align}
where the scheme for $\alpha_s$ does not need to be specified at this order.  

\section{Expansion for $x \rightarrow 1$}\label{sec:expansion}

\begin{figure*}
 \centering

  \subfigure{\epsfig{file=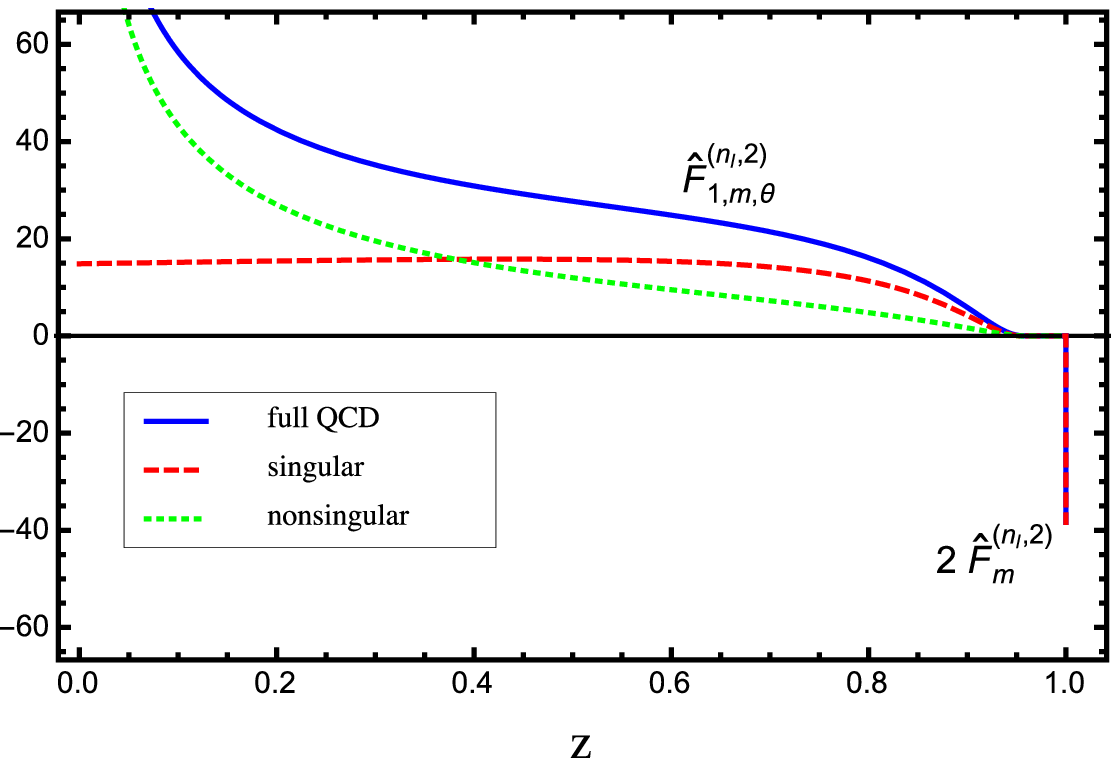,width=0.45\linewidth,clip=}}\hfill
  \subfigure{\epsfig{file=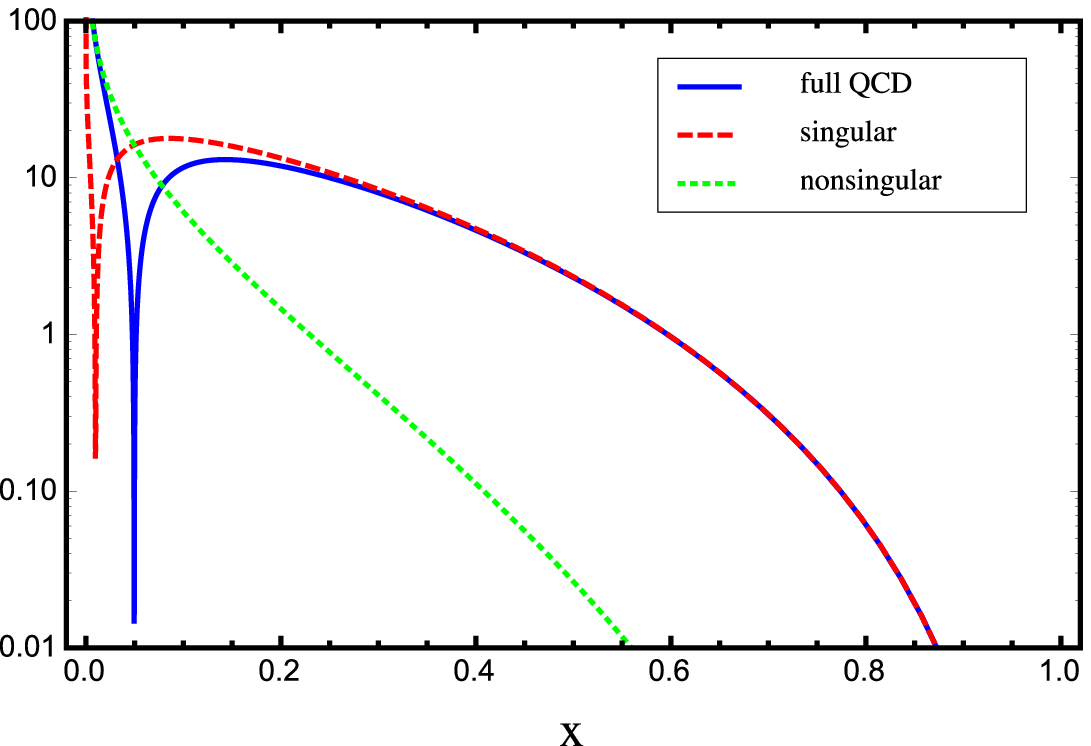,width=0.45\linewidth,clip=}}\hfill
 \caption{Secondary massive quark contributions at $\mathcal{O}(\alpha_s^2 C_F T_F)$ for $\hat{m}=m/Q=0.1$ in full QCD (blue, solid) together with the singular result for $z\rightarrow 1$ at fixed order (red, dashed) and the nonsingular terms (green, dotted), all normalized by $\alpha_s^2 C_F T_F/(4\pi)^2$. The left panel shows the purely partonic result with initial state quarks, while in the right panel we convoluted the partonic form factors with the function $(1-x)^4$ representing the steep decrease of the PDFs for large values of $x$. \label{fig:DIS_massivequark}}  
\end{figure*}

The massive quark corrections to the factorization theorem discussed in Sec.~\ref{sec:massive} represent the singular $\mathcal{O}(\alpha_s^2 C_F T_F)$ secondary massive quark corrections to the structure function $F_1(x,Q,m)$ in the fixed-order expansion in full QCD. Besides the virtual contributions in QCD, which are fully contained in the SCET description, the singular perturbative fixed-order corrections also consist of the collinear real radiation contributions which arise for $1-x\sim m^2/Q^2\ll 1$.  Setting $\mu=\mu_H=\mu_J=\mu_m$ in Eq.~\eqref{eq:facttheomassive} we obtain
  \begin{align}
 & \left.F_1 (x,Q,m)\right|_{\rm FO}=  \sum\limits_{i=q} \frac{e_i^2}{2}\int \!\df \xi \, \phi_{i/P}^{(n_l)}\Big(\xi-x,\mu\Big) \nn \\& \times \underbrace{Q^2 H^{(n_l)}(Q,m,\mu) \, J^{(n_l)}(Q^2(1-\xi),m,\mu)}_{\equiv \,H^{(n_l)}_{qq,z \to 1}(\xi,Q,m,\mu)}  \, . 
\end{align}
where the massive quark contributions to the fixed-order hard function at $\mathcal{O}(\alpha_s^2 C_F T_F)$ read
\begin{align}\label{eq:Csing_m}
 H^{(n_l,2)}_{qq,z \to 1,m}(z,Q,m) = & \, 2 \hat{F}^{(n_l,2)}_{m}(Q,m) \, \delta(1-z)\nn\\&+ Q^2 J^{(n_l,2)}_{m, \rm real}(Q^2(1-z),m) \, .
\end{align}
with $\hat{F}^{(n_l,2)}_{m}(Q,m)$ and $J_{m,\rm real}^{(n_l,2)}(s,m)$ given in Eqs.~(\ref{eq:F_QCD}) and~(\ref{eq:J_real}) with $\alpha_s=\alpha_s^{(n_l)}(\mu)$. We can obtain this result also from the corresponding full QCD fixed-order result in Appendix~\ref{sec:fixed-order}. Since the virtual contributions multiplied by the $\delta(1-z)$ distribution  in the OPE and endpoint regions agree, we only have to consider the expansion of the real radiation term $\hat{F}^{(n_l,2)}_{1,m,\theta}(z,Q,m)$ in Eq.~(\ref{eq:f_QCD_theta}) for $1-z\sim \hat{m}^2\ll 1$ (with $\hat{m}=m/Q$), which yields indeed the correct term,
\begin{align}
&\theta(1-z-4 \hat{m}^2 z) \, \hat{F}^{(n_l,2)}_{1,m, \theta}(z,Q,m) \\&\qquad\stackrel{z\rightarrow 1}{\longrightarrow}  \,\, Q^2 J_{m,\rm real}^{(n_l,2)}(Q^2(1-z),m) + \,\mathcal{O}((1-z)^0,\hat{m}^0) \,. \nn
\end{align}
In Fig.~\ref{fig:DIS_massivequark} we investigate how well the expansions work for the specific scale ratio $\hat{m}=0.1$. The left panel shows the partonic result for the full QCD corrections (blue, solid), i.e.~$\hat{F}^{(n_l,2)}_1 (z,Q,m)$ in Eq.~(\ref{eq:Fnl_tot}), the singular result for the endpoint region at fixed order (red, dashed), i.e.~$H^{(n_l,2)}_{qq,z \to 1,m}(z,Q,m) $ in Eq.~(\ref{eq:Csing_m}), and the difference describing the nonsingular corrections (green, dotted). We see that for $1-z \lesssim 4m^2/Q^2$ the endpoint corrections encode the dominant behavior, but fail to give a good description of the full QCD form factor below a certain value, here $z \lesssim 0.5$. The right panel displays the absolute value of the convolution between these partonic functions and a common function $f(x) =(1-x)^4$ acting as a dummy PDF at the endpoint which falls off steeply for $x \to 1$. The convoluted results for the full QCD and singular terms are negative for $x \gtrsim 0.05$. We see that at this 
level the agreement between the singular and full results is much better due to dynamical threshold enhancement (see e.g.~Ref.\cite{Becher:2007ty}), up to values significantly lower than $x = 0.5$. This may have the consequence that endpoint region effects can have an impact even at smaller values of $x$ probed at hadron-hadron colliders. A recent analysis on this issue was carried out in Ref.~\cite{Bonvini:2015ira}. They found that the effect can be sizable and may require the use of resummed PDFs for resummed calculations.

\onecolumngrid
\vspace{\columnsep}
\vspace{\columnsep}
\twocolumngrid

\bibliography{DIS}{}
\bibliographystyle{my_bibstyle}
 
\end{document}